\documentclass[twocolumn]{aastex62}
\usepackage{amsfonts,amsmath,amssymb,amsthm,epsfig,graphicx,float,tabularx,enumitem}
\begin{document}

\title{Wave-driven outbursts and variability of low-mass supernova progenitors}

\correspondingauthor{Samantha Wu}
\email{scwu@astro.caltech.edu }

\author{Samantha C. Wu}
\affiliation{California Institute of Technology, Astronomy Department, Pasadena, CA 91125, USA}

\author{Jim Fuller}
\affiliation{TAPIR, Mailcode 350-17, California Institute of Technology, Pasadena, CA 91125, USA}

\begin{abstract}
In a substantial number of core-collapse supernovae, early-time interaction indicates a dense circumstellar medium (CSM) that may be produced by
outbursts from the progenitor star. Wave-driven mass loss is a possible mechanism to produce these signatures, with previous work suggesting this mechanism is most effective for low mass ($ \sim \! 11 \, M_\odot$) SN progenitors. 
Using one-dimensional hydrodynamic simulations with MESA, we study the effects of this wave heating in 
SN progenitors of masses~$M_{\rm ZAMS} = 10\text{--}13\, M_{\odot}$. This range encompasses stars that experience semi-degenerate, central neon burning and more degenerate, off-center neon ignition. We find that central Ne ignition at~$M_{\rm ZAMS} = 11\, M_{\odot}$ produces a burst of intense wave heating that transmits~$\sim10^{47}$ erg of energy at~$10$ years before core collapse, whereas other masses experience smaller levels of wave heating.
Wave heating does not hydrodynamically drive mass loss in any of our models and is unlikely to produce a very massive CSM on its own. However,
wave heating can cause large radial expansion (by more than an order of magnitude), photospheric cooling, and luminosity brightening by up to~$\sim 10^6\, L_{\odot}$ in hydrogen-poor stripped star models. Some type Ib/c progenitors could drastically change their appearance in the final years of their lives, with brightness in visual bands increasing by nearly 3 mags. Moreover, interaction with a close binary companion could drive intense mass loss, with implications for type Ibn and other interaction-powered SNe.
\end{abstract}

\section{Introduction}

A subset of core-collapse supernovae (SNe) across nearly all spectroscopic classes exhibit signatures of early interaction with circumstellar material (CSM). In such SNe, the CSM interaction is observed through fast-rising, bright early-time light curves; featureless, blue early spectra that signify a shock-heated envelope or CSM; or narrow emission lines in early spectra that appear when nearby CSM is flash-ionized by the SN breakout. A few recent examples of SNe that demonstrate evidence for early interaction with CSM include the type II events SN2018zd \citep{ZhangJ2020}, SN2013fs \citep{Yaron2017}, and several type II events detected by ZTF \citep{bruch2021}, as well as the type I events LSQ13ddu \citep{Clark2020}, SN2018gep \citep{Ho2019}, SN2019dge \citep{Yao2020}, SN2019uo \citep{Gangopadhyay2020}, and SN2019yvr \citep{Kilpatrick2021}. The CSM observed in these events is usually attributed to elevated mass loss from the progenitor stars that precedes the SN. For type II SNe with early peaks in the light curve, the CSM signatures may also be explained by roughly one solar mass of material in an optically thin stellar chromosphere or corona \citep{Dessart2017,Hillier2019}. 

Other events present late-time interaction and narrow emission lines, indicating mass loss occurring decades before the SN. Some recent examples include SN2017dio \citep{Kuncarayakti2018}, SN2017ens \citep{Chen2018}, SN2013L \citep{Taddia2020}, iPTF16eh \citep{Lunnan2018}, and SN2004dk \citep{Mauerhan2018,Pooley2019}.
Many other events with both early- and late-time interaction are listed in \citet{Wu2021}, \citet{fuller2018}, and \citet{fuller2017}.

While pre-SN outbursts have been observed directly in several SN progenitors \citep{Ofek2014}, the majority of SN progenitors may not experience pre-SN outbursts or amplified variability at all. \citet{Kochanek2016} found that progenitor constraints on nearby, well-studied SNe do not support evidence for outbursts, and \citet{Samson2017} found no significant variability in the progenitor for type Ic SN2012fh. \citet{Johnson2018} constrained variability amplitudes to less than~$\sim 10\%$ in four type II-P progenitors. As a result, the mechanisms behind pre-SN mass loss must operate such that they manifest a range in behavior, from elevated mass loss and bright pre-SN outbursts to no unusual variations at all.

One promising explanation for pre-SN outbursts is the wave heating model proposed by \citet{Quataert2012}. Internal gravity waves (IGW) are excited as a routine consequence of convection, and the vigorous convection that occurs during late burning stages in massive stars may generate waves carrying more than~$10^7\, L_{\odot}$ of power. These IGW couple with acoustic waves to deposit some of their power in the outer layers of the star, which may be able eject mass or drive outbursts. \citet{Shiode2014} found that more massive stars produced larger wave heating rates, but ensuing outbursts occurred closer to core-collapse. Modeling the wave heating in a~$15\, M_{\odot}$ red supergiant, \citet{fuller2017} found that waves could inflate the envelope and drive a mild outburst, potentially causing mass loss through a secondary shock. \citet{fuller2018} examined hydrogen-poor stars in which they found that wave heating could launch a dense, super-Eddington wind of~$\dot{M} \sim 10^{-2}\, M_{\odot}/\text{yr}$. These studies also predicted large variations in luminosity and temperature as a result of wave heating.

In \citet{Wu2021}, we updated the physical model used for wave generation and propagation in \citet{fuller2017} and \citet{fuller2018} to include non-linear wave breaking effects and to more realistically model the wave spectrum excited by convection. By simultaneously accounting for these effects throughout the evolution of core convective shells in a suite of stars with~$M_{\rm ZAMS} = 11 \text{--} 50\, M_{\odot}$, \citet{Wu2021} found that wave heating rates were an order of magnitude lower than adopted in prior work. While most models were unlikely to produce observable pre-SN outbursts, the lowest- and highest-mass stars experienced the highest levels of wave heating and were favored to drive outbursts. \citet{leung2021} surveyed the upper end of this mass range with the updated physics, extending up to~$M_{\rm ZAMS} = 70\, M_{\odot}$ and including hydrogen-poor stars, and likewise predicted less energy deposited in the envelope and less mass loss than prior work. Their hydrodynamical simulations of the most energetic high-mass model ejected at most~$0.01\, M_{\odot}$ of material close to core collapse, forming quite confined CSM.

In this paper, we study the low-mass end of models in \citet{Wu2021}, where outbursts are promising due to semi-degenerate central neon ignition, and we now deposit the wave heat in our models as they evolve. We extend our scope down to~$M = 10\, M_{\odot}$ stars to capture the physics of waves generated by semi-degenerate, off-center neon ignition.
Additionally, we apply the same wave heating physics to hydrogen-poor models for stripped progenitors. In all our models, we add the wave heat and explore their hydrodynamical response. We find that our updated wave heating rates are not able to drive significant mass loss in either supergiant or stripped-star models and that wave heating only causes small-amplitude surface variability in our supergiant models. However, in our stripped star models we predict large outbursts in luminosity as well as large changes in the photospheric radius and surface temperature, which may be detectable in progenitors of type Ib SNe.

\section{Implementation of Wave Physics in Stellar Models}
\label{sec:physics}

\subsection{Wave Generation and Propagation}
\label{sec:equations1}

To implement wave energy transport, we follow the procedure of \citet{Wu2021}. We summarize the physics of wave generation and propagation here and refer the reader to Section 2 of \citet{Wu2021} for more details.

Gravity waves are excited at the interface between convective and radiative zones and carry a fraction of the kinetic energy of turbulent convection. While our understanding of this process remains incomplete, the power put into waves,~$L_{\rm wave}$, is at least \citep{Lecoanet2013}
\begin{equation}
\label{eq:lwave}
    L_{\rm wave} = \mathcal{M}_{\rm con} L_{\rm con}
\end{equation}
where~$\mathcal{M}_{\rm con}$ is the MLT convective Mach number and~$L_{\rm con}$ is the convective luminosity \citep{Goldreich1990}. We define the convective velocity ~$v_{\rm con}$ as
\begin{equation}
\label{eq:Lcon}
    v_{\rm con} = \left(\frac{ L_{\rm con}}{4\pi\rho r^2}\right)^{1/3} \, ,
\end{equation}
and we define the associated MLT convective turnover frequency as
\begin{equation}
\label{eq:omega}
    \omega_{\rm con} = 2\pi \frac{v_{\rm con}}{2 \alpha_{\rm MLT} H},
\end{equation}
where~$\alpha_{\rm MLT} H$ is the mixing length and~$H$ is the scale height.

As in \citet{Wu2021}, we use the power spectrum of waves over angular wavenumber~$\ell$ from \citet{Goldreich1990} and \citet{Shiode2013} to calculate the proportion of wave energy generation per~$\ell$ value,~$\dot{E}_\ell/L_{\rm wave}$:
\begin{align}
\label{eq:ellpowerspectrum}
    \frac{d\dot{E}_g}{d \ln \omega d \ln \ell} &\sim  
    L_{\rm wave} \left( \frac{\omega}{\omega_{\rm{con}}}\right)^{\!-a}
    \left( \frac{\ell}{\ell_{\rm con}} \right)^{\!b+1}
    \left(1+\frac{\ell}{\ell_{\rm con}}\right) \nonumber \\
    & \times \exp \left[-\left(\frac{\ell}{\ell_{\rm con}}\right)^{\!2} \left(\frac{\omega}{\omega_{\rm con}}\right)^{\!-3} \right] \, .
\end{align}
Here,~$\ell_{\rm con} = r/\min(H,\Delta r)$ is evaluated at the edge of the convective zone and the predicted exponents are $a=13/2$ and $b=2$. This spectrum peaks near the value of~$\ell_{\rm con}$ (see Figure 1 in \citealt{Wu2021}). As Equation \ref{eq:ellpowerspectrum} shows, the wave power drops off steeply for~$\omega>\omega_{\rm con}$, and we do not expect low-frequency waves with $\omega<\omega_{\rm con}$ to contribute much wave power due to tunneling and damping effects described below. We therefore simplify the calculation by setting~$\omega = \omega_{\rm con}$ for all~$\ell$ values, and we normalize the power spectrum to~$L_{\rm wave}$ so that~$\sum \dot{E}_\ell = L_{\rm wave}$.

In the WKB limit, linear waves have the dispersion relation
\begin{equation}
\label{eq:dispersion}
    k_r^2=\frac{(N^2-\omega^2)(L_{\ell}^2-\omega^2)}{\omega^2 c_s^2},
\end{equation}
where~$N^2$ is the squared Brunt-V\"ais\"al\"a frequency,~$k_r$ is the radial wavenumber, and~$L_{\ell}^2 = {\ell}({\ell}+1)c_s^2/r^2$ is the Lamb frequency squared. In the limit that~$\omega \ll N, L_{\ell}$, this reduces to the gravity wave dispersion relation
\begin{equation}
\label{eq:gravitykr}
    k_r^2=\frac{\ell(\ell+1)N^2}{\omega^2 r^2}
\end{equation}
with group velocity 
\begin{equation}
\label{eq:gravityvg}
    v_g=\omega^2 r/\sqrt{\ell(\ell+1)N^2}.
\end{equation}
The limit~$\omega \gg N, L_{\ell}$ gives acoustic waves, with dispersion relation
\begin{equation}
    k_r^2=\frac{\omega^2}{c_s^2}
\end{equation}
and group velocity~$v_g=c_s$. In either of these limits, linear waves propagate freely and approximately conserve their luminosity, apart from damping effects discussed below.

If~$\omega > N$ and~$\omega < L_{\ell}$ or vice versa, then the radial wavenumber is imaginary and waves are evanescent. The probability of tunneling through this evanescent zone, or the fraction of transmitted wave energy, is approximately given by the transmission coefficient
\begin{equation}
\label{eq:transmissioncoeff}
    T^2 = \exp\left(-2 \int_{r_0}^{r_1}  \lvert k_r \rvert dr \right),
\end{equation}
where the integral is taken over the evanescent zone. As the thickest evanescent zone dominates the wave reflection (see Appendix B2 of \citealt{fuller2017}), we take the minimum value of~$T^2$ out of all evanescent zones,~$T_{\rm min}^2$, to calculate the wave flux tunneling into the envelope. The remaining fraction~$1-T_{\rm min}^2$ of wave energy that encounters the evanescent region is reflected from the boundary of the evanescent zone.

In addition, after traveling to the upper edge of the core and back, a wave's energy is attenuated by the factor
\begin{equation}
\label{eq:fnu}
    f_{\nu} = \exp\left[2\int_{r_{-}}^{r_{+}} \frac{\gamma_{\nu} + \gamma_{\rm rad}}{v_g} dr\right]
\end{equation}
where~$\gamma_\nu$ and~$\gamma_{\rm rad}$ are the neutrino and thermal wave energy damping rates respectively,~$v_g$ is the gravity wave group velocity (Equation~\ref{eq:gravityvg}), and the integral is taken over the upper and lower boundaries of the gravity wave cavity. \textbf{Note that the definition of $f_\nu$ assumes the weak damping limit, where the majority of wave energy escapes only after traversing the core multiple times; in the limit of strong damping, where most of the wave energy escapes after traveling to the upper edge of the core once, it is appropriate to use $\sqrt{f_\nu}$ instead. We find that our models are almost always in the weak damping limit, especially when significant wave heating occurs, justifying our approximation for $f_\nu$.}

The fraction of wave energy that escapes the core,~$f_{\rm{esc}, \ell}$, is then determined by the transmission coefficient (Equation~\ref{eq:transmissioncoeff}) and energy losses within the core,
\begin{equation}
\label{eq:heatfrac}
    f_{\rm esc, \ell} = \left(1+\frac{f_{\nu}-1}{T_{\rm min}^2} \right)^{-1}.
\end{equation}
Given the escape fraction, the~$\ell$-dependent power that escapes to heat the envelope is
\begin{equation}
\label{eq:lheatell}
    L_{\rm heat, \ell} = f_{\rm esc, \ell} \dot{E}_\ell.
\end{equation}

Equation~\ref{eq:lheatell} holds for waves that remain linear, but waves that experience non-linear wave breaking will experience another source of energy loss before escaping to heat the envelope. A measure of the gravity wave non-linearity as a function of~$\ell$ is 
\begin{equation}
\label{eq:nonlinearity}
    \left|k_r \xi_r\right| = \left[\frac{2}{T_{\rm min}^2}\frac{f_{\rm esc, \ell} \dot{E}_\ell N [\ell (\ell+1)]^{3/2}}{4\pi\rho r^5\omega^4} \right]^{1/2}.
\end{equation} 
Where~$\left|k_r \xi_r\right| \geq 1$, waves are non-linear, whereas linear waves have~$\left|k_r \xi_r\right| \leq 1$. Such non-linear waves will break and experience an energy cascade to small scales, where their energy dissipates \citep{Barker2010}. This effectively caps the wave amplitude~$\xi_r$ so that~$|k_r\xi_r| \lesssim 1$, therefore reducing the wave power by a factor~$|k_r\xi_r|^2$ when~$|k_r\xi_r| \geq 1$. Thus for non-linear waves, the power that escapes to heat the envelope is effectively
\begin{align}
\label{eq:lheatellnl}
    L_{\rm heat, \ell} &= f_{\rm esc, \ell} \dot{E}_\ell/|k_r\xi_r|^2 \nonumber \\
    &= \frac{T_{\rm min}^2}{2} \frac{4\pi\rho r^5\omega^4}{N [\ell (\ell+1)]^{3/2}} \, .
\end{align}
We use the maximum value of~$|k_r\xi_r|$ (i.e., the minimum value of the second line in equation \ref{eq:lheatellnl}) in the g-mode cavity to compute the above reduction of wave power due to non-linear damping.

\subsection{Wave dissipation}
\label{sec:equations2}

To model the wave damping and energy deposition in the envelope, we follow the methods of \citet{fuller2018}, accounting for background flows as explained in their Section 3.4. We summarize the approach here.

Wave dissipation in the envelope can be understood by considering the damping mass, which is the mass through which a wave must pass to dissipate all its energy, given by
\begin{equation}
    M_{\rm damp} = L_{\rm{ac}} \left( \frac{d L_{\rm{heat,co}}}{dm}\right)^{-1}.
\end{equation}
in the presence of flows. Here~$d L_{\rm{heat,co}}/dm$ is the wave heating rate per unit mass and $L_{\rm{heat,co}}$ is the wave energy flux measured in the comoving frame. The conserved quantity (in the absence of damping) in the inertial frame is the wave action $L_{\rm ac} = 2\pi r^2 \rho c_s u^2 (1+v_r/c_s)$, where~$u$ is the radial velocity amplitude of the wave.

\citet{fuller2018} find that the damping mass due to radiative diffusion is  
\begin{equation}
    M_{\rm damp, rad} = \frac{2L_{\rm max}}{\omega^2 K} \left(1+\frac{v_r}{c_s}\right)^2   
    \label{eq:Mdamprad}
\end{equation}
where the maximum possible wave flux in the linear regime is~$L_{\rm max} = 2\pi r^2 \rho c_s^3$. \citet{fuller2018} find that the damping mass due to weak shock dissipation is
\begin{equation}
\begin{split}
\label{eq:Mdampshock}
    M_{\rm damp, shock} & = \frac{3\pi}{\gamma + 1}\frac{L_{\rm max}}{\omega c_s^2} \\
    & \times \left(\frac{L_{\rm max}\left(1 + v_r/c_s\right)^5}{L_{\rm ac}}\right)^{1/2}.
\end{split}
\end{equation}
Thus the effective damping mass is
\begin{equation}
    M_{\rm damp} = \left[ M_{\rm damp,shock}^{-1} + M_{\rm damp, rad}^{-1} \right]^{-1},
    \label{eq:Mdamp}
\end{equation}
so that the energy deposited per unit time per unit mass in our models is
\begin{equation}
\label{eq:dLdm}
    \epsilon_{\rm wave}  = \frac{d L_{\rm{heat,co}}}{dm} = \frac{L_{\rm ac}}{M_{\rm damp}}.
\end{equation}
Since $L_{\rm ac}$ is the quantity conserved in the absence of damping, we compute the energy deposited from escaping waves (with initial $L_{\rm ac} =  L_{\rm{heat},\ell}$) as
\begin{equation}
\label{eq:dLdmactual}
\epsilon_{\rm wave}=L_{\rm{heat},\ell}/M_{\rm damp}.
\end{equation}

\begin{figure}
    \centering
    \includegraphics[width=\columnwidth]{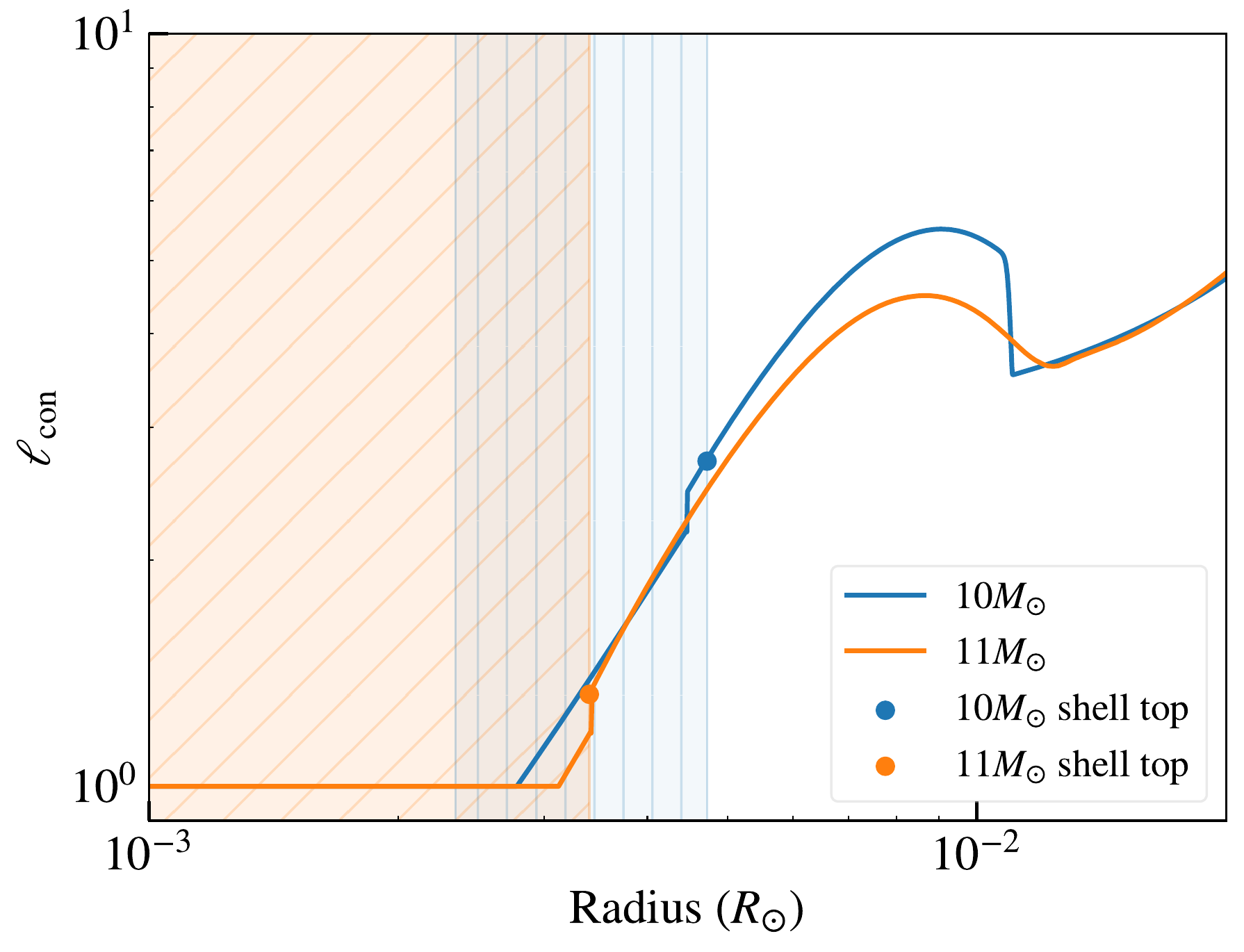}
    \caption{The value of~$\ell_{\rm con} = r/\min(H,\Delta r)$ as a function of radius for the Ne burning convective region in the~$10\, M_{\odot}$ (blue) and~$11\, M_{\odot}$ (orange) supergiant models. The radial extent of the~$10\, M_{\odot}$ burning shell is indicated by the blue shaded vertically hatched region, and the radial extent of the~$11\, M_{\odot}$ core burning is shaded in orange with diagonal hatching. The radius of the top of the burning region, where~$\ell_{\rm con}$ is evaluated, is marked with a circle for each model. Since the~$10\, M_{\odot}$ model burns Ne off-center, the top of its Ne burning shell occurs at larger radius, increasing its~$\ell_{\rm con}$ value. This also applies to the stripped star models.}
    \label{fig:10v11lcon}
\end{figure}

\begin{figure*}
    \centering
    \includegraphics[width=\textwidth]{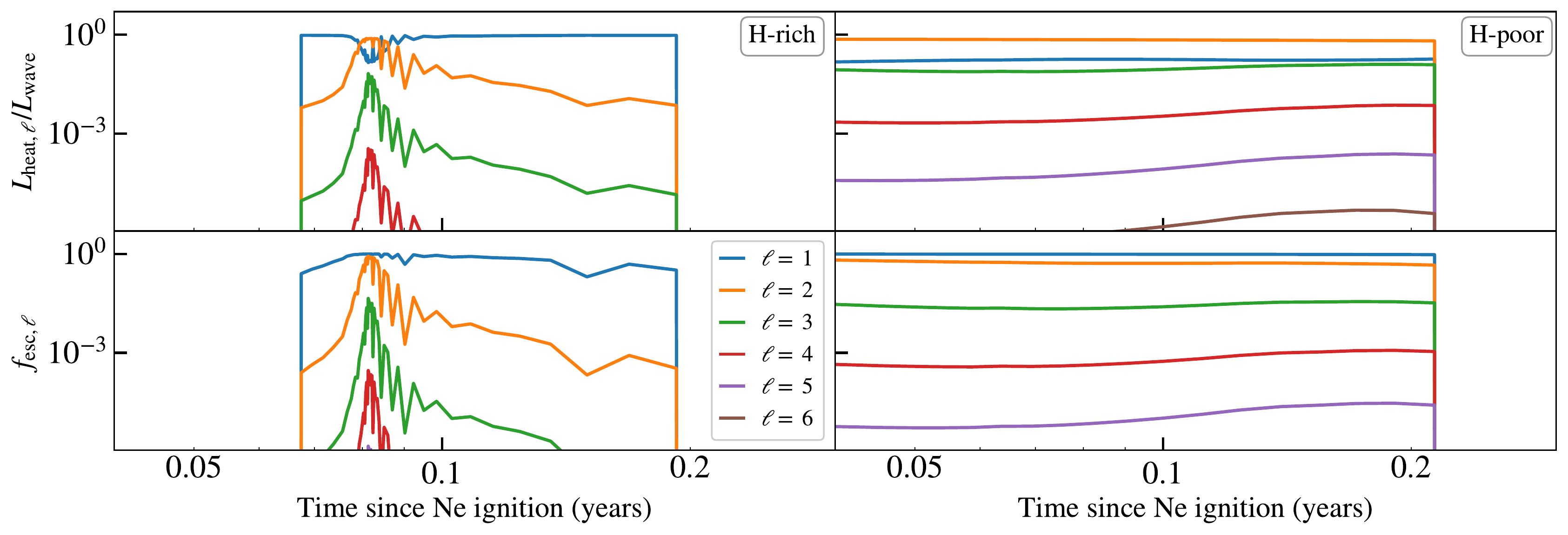}
    \caption{Fraction of envelope wave heating caused by each angular number~$\ell$ (top), and the wave escape probability for each~$\ell$ (bottom), as a function of time since Ne ignition in the~$M_{\rm ZAMS} = 10\, M_{\odot}$ models. The left figure shows the supergiant model and the right figure shows the stripped star model. Results are only shown for waves where~$L_{\rm heat,\ell} > 10^3\, L_{\odot}$. }
    \label{fig:10fesc}
\end{figure*}

\begin{figure}
    \includegraphics[width=\columnwidth]{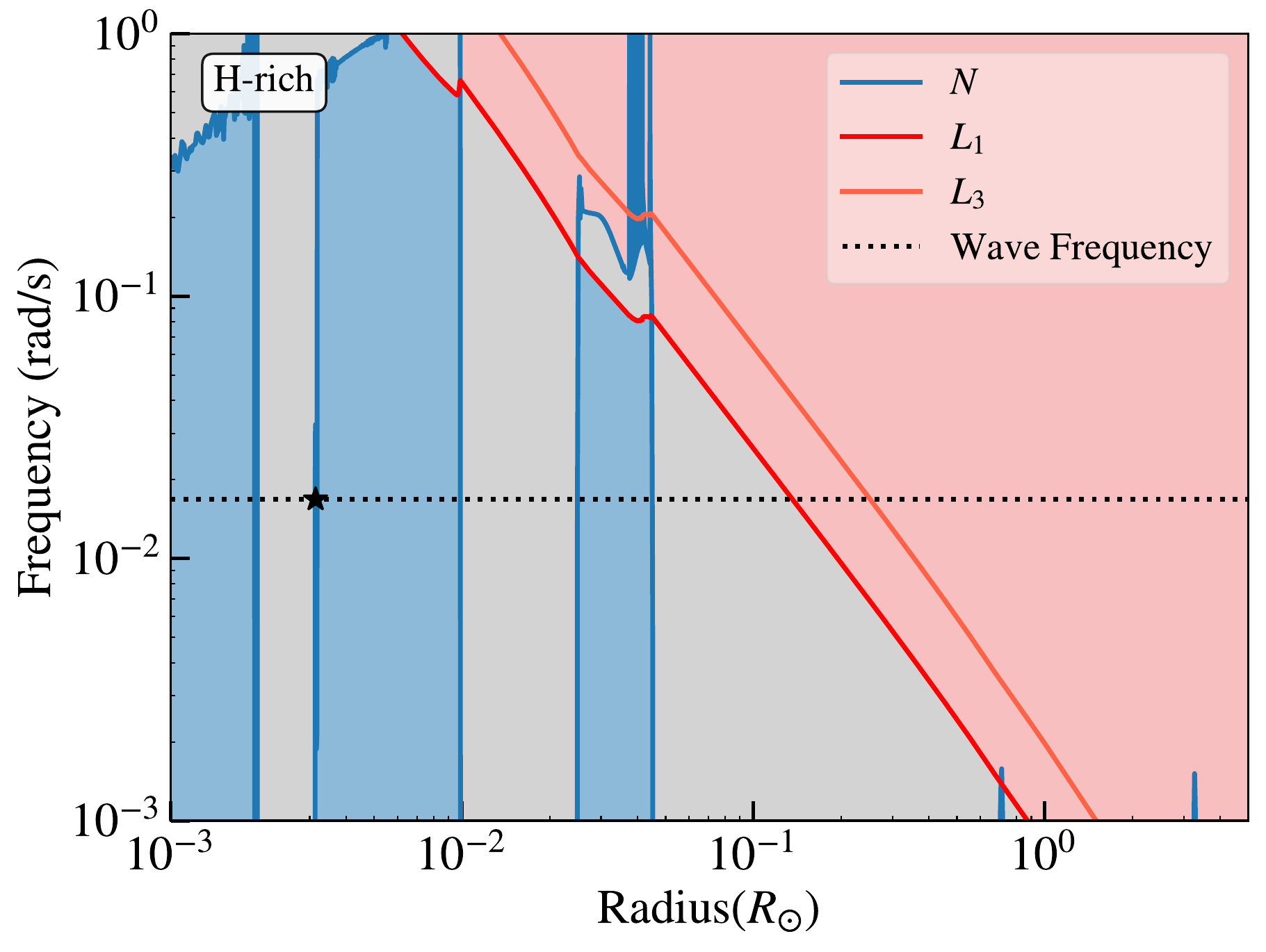}
    \includegraphics[width=\columnwidth]{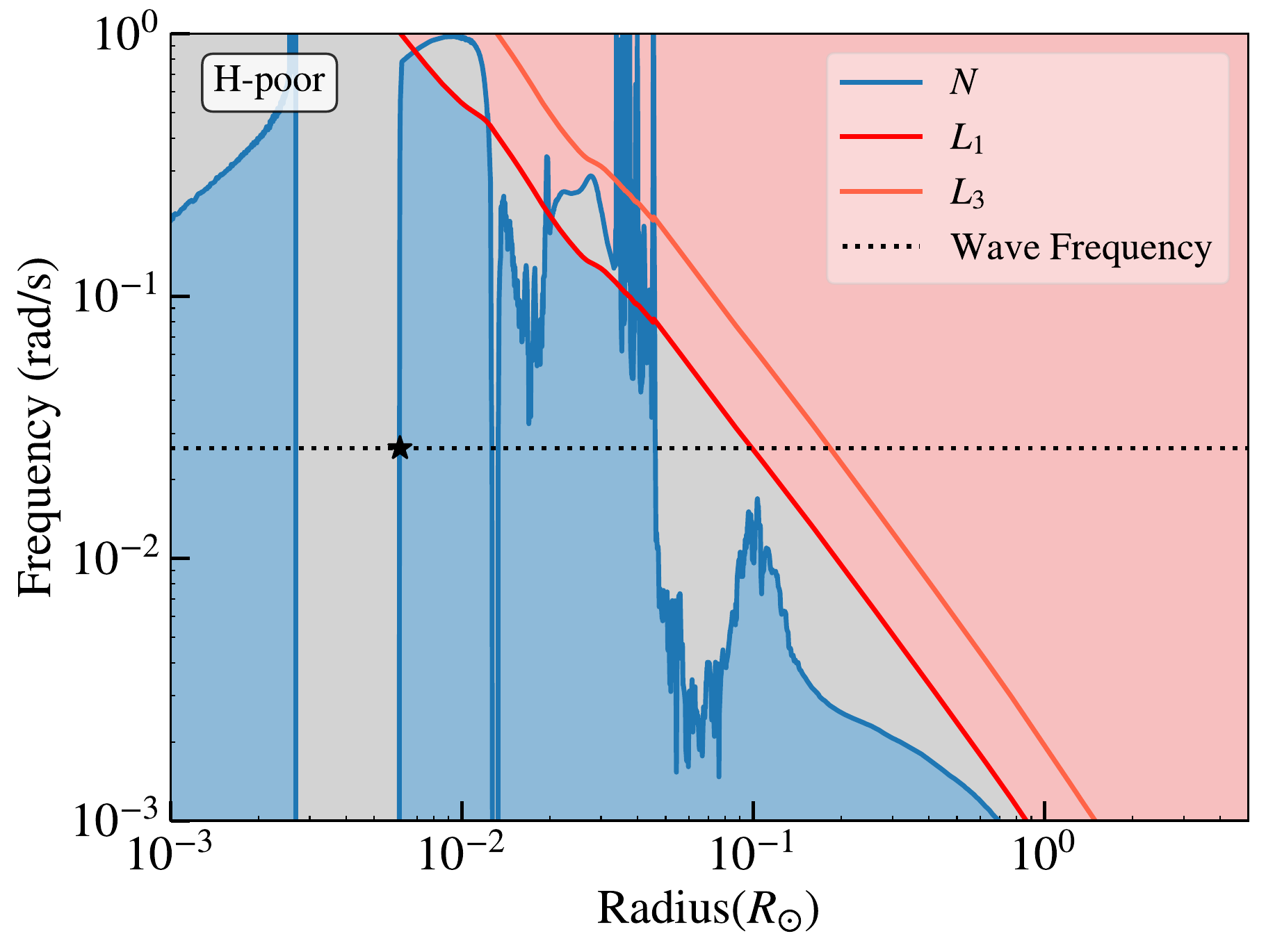}
    \caption{Propagation diagrams for the~$M_{\rm ZAMS} = 10\,  M_{\odot}$ supergiant model (top) and stripped star model (bottom), each during Ne burning. Shown are the Brunt-V\"ais\"al\"a frequency N (blue) and the Lamb frequency~$L_{\ell}$ for~$\ell = 1,3$ (shades of red). Waves propagate through the gravity wave cavity (blue region) and into the envelope as acoustic waves (red region, shaded for~$\ell = 1$), tunneling through evanescent zones (gray region) along the way. Off-center Ne ignition in the core excites waves with~$\omega \sim \! 10^{-2}$ rad/s (dotted line) at the top of the convective zone (star). The stripped star model has higher~$\omega$, which allows the waves to tunnel through a thinner evanescent zone and causes the escape fraction to be higher for waves of all~$\ell$ (Figure~\ref{fig:10fesc}.)
    }
    \label{fig:10Mprop}
\end{figure}

\begin{figure*}
    \centering
    \includegraphics[width=0.494\textwidth]{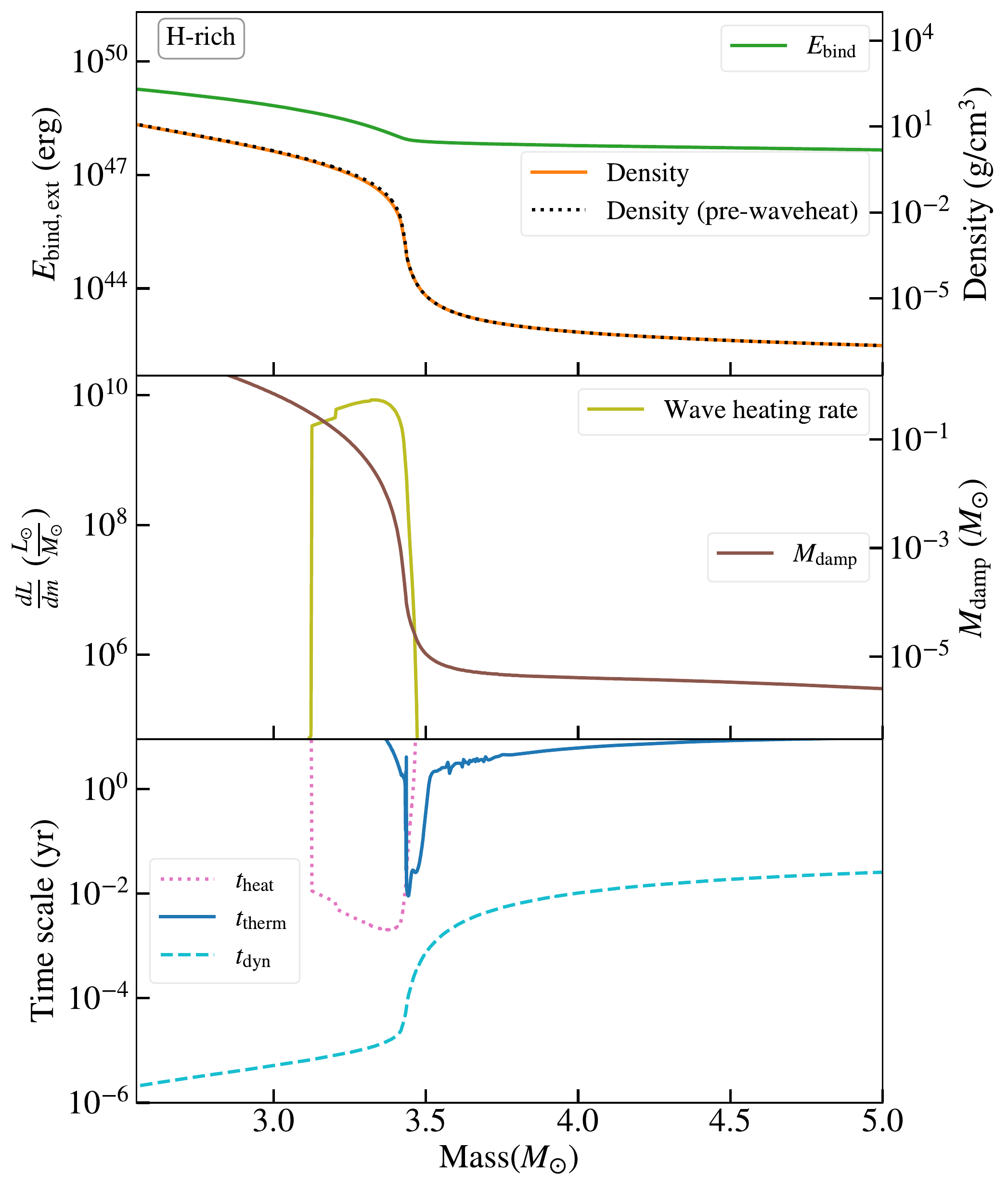}
    \includegraphics[width=0.4935\textwidth]{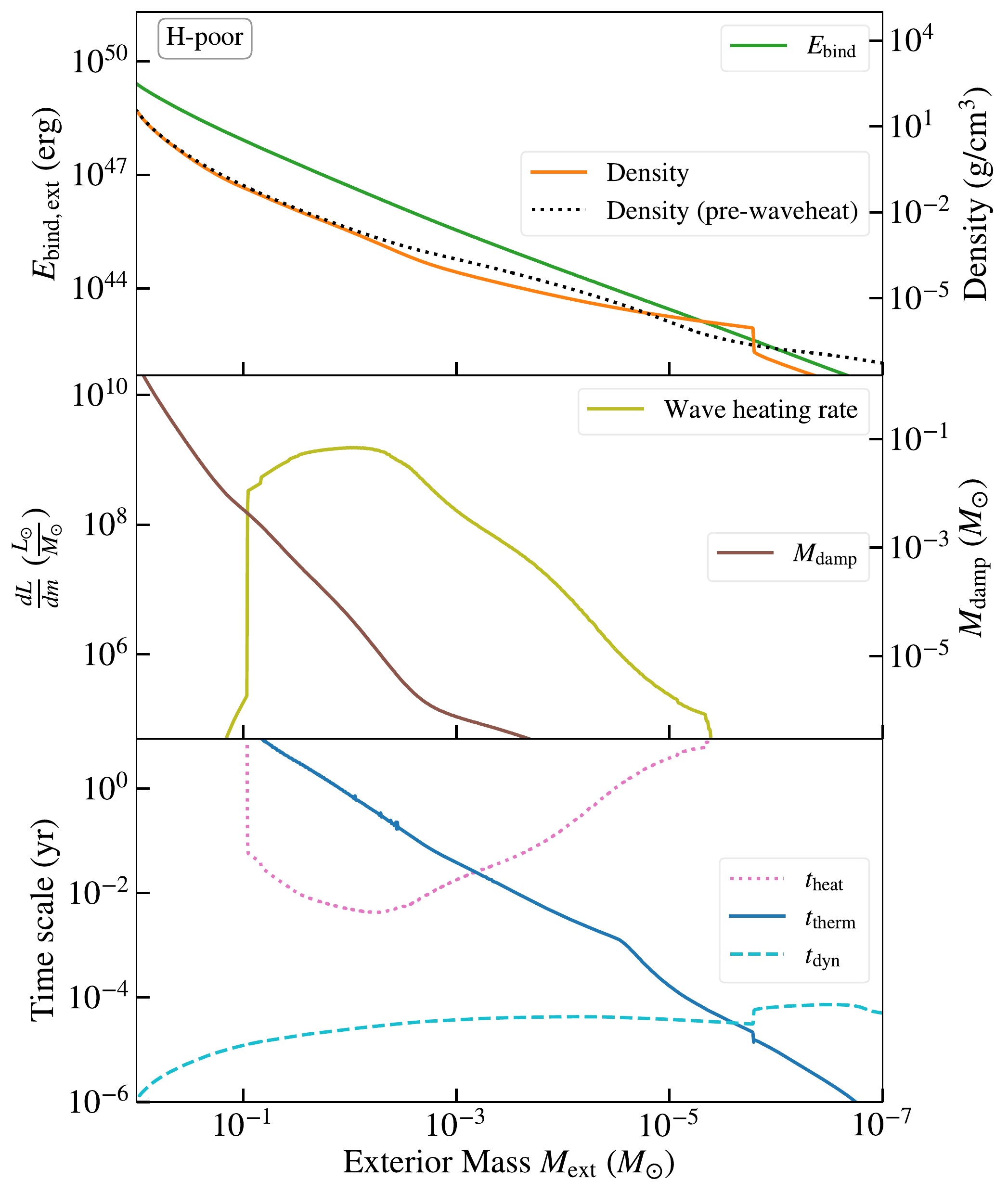}
    
    \caption{Left panel: Wave heating diagnostics for an~$11\, M_{\odot}$ red supergiant model during Ne burning as a function of mass coordinate. Top: Binding energy integrated inward from the surface of the model and the star's density profile, before and after wave heating due to Ne burning. Middle: Wave energy deposition rate per unit mass and the damping mass (i.e., $4 \pi \rho r^2$ times the damping length, Equation~\ref{eq:Mdamp}). Bottom: Wave heating timescale ($t_{\rm heat}$), local thermal timescale ($t_{\rm therm}$), and local dynamical timescale ($t_{\rm dyn}$). Right panel: Same as left for the ~$M_{\rm ZAMS}=11\, M_{\odot}$ stripped star during Ne burning, but as a function of exterior mass coordinate. 
    }
    \label{fig:11Ebind}
\end{figure*}

\subsection{Methods}
\label{sec:methods}
We run a suite of MESA simulations \citep{mesa2011,mesa2013,mesa2015,mesa2018,mesa2019} for zero-age main sequence (ZAMS) masses 10, 10.5, 11, 12, and 13~$M_{\odot}$ and evolve the stars from the main sequence to at least five years after neon (Ne) ignition. For the~$11\text{--}13\, M_{\odot}$ models, we evolve the stars until core collapse; however, the~$10$ and~$10.5\, M_{\odot}$ models experience off-center Ne burning that makes these stars prohibitively computationally expensive to run until core collapse, so we simulate these models through Ne ignition and at least five years of Ne burning flame propagation. In addition, we create hydrogen-poor (H-poor) stellar models following the methods of \citet{fuller2018}, Appendix A\footnote{Model parameters are available at https://zenodo.org/communities/mesa.}. Starting with models of the same initial masses ($M_{\rm ZAMS}$) as the aforementioned hydrogen-rich (H-rich) supergiant models, we remove the hydrogen envelope after core helium (He) burning so that the resulting H-poor, stripped star models have nearly the same He core masses as the H-rich models. Accordingly, the stripped star models with initial mass~$10$ and~$10.5\, M_{\odot}$ experience the same off-center Ne ignition as their supergiant counterparts, while all other stripped star models are evolved until core collapse.


Beginning at core carbon burning, we capture the wave-driven mass loss as in \citet{fuller2018} by setting the outer boundary at an optical depth of~$\tau = 10^{-2}$. Material that flows past this outer boundary is removed from the grid, and we define mass loss rates in the paper as the mass flux through this outer boundary. At each timestep, we perform the same calculations described in Section 2.2 of \citet{Wu2021} and above for the convective burning regions in each model. This gives us the amount of wave heat transmitted, including possible non-linear effects. We then add the wave heat in each cell of the model using Equation~\ref{eq:dLdm}.

\section{Results}
\label{sec:results}

\subsection{Wave Heating Rates}
\label{sec:results1}
In all our models, the most important contribution to wave heating occurs during Ne burning roughly $5-20$ years before core collapse. The next major contribution to wave heating from Si burning is energetic, but occurs too near core collapse for waves excited by Si burning to reach the surface of our supergiant progenitors. We do not simulate Si burning in our~$M < 11\, M_{\odot}$ models.

The~$M < 11\, M_{\odot}$ models in this work extend our wave heating study down to lower-mass stars in which Ne ignites off-center. This difference greatly affects the amount of wave power produced by the~$10 \text{--} 10.5\, M_{\odot}$ supergiant models, bringing the wave energy deposition rate down by~$1\text{--}2$ orders of magnitude compared to the less degenerate and centrally burning~$11\, M_{\odot}$ model (see Figure~\ref{fig:10to13surf}, top left panel).
The main reduction in wave heating arises from lower wave escape fractions from the core occurs because the excited wave spectrum peaks at higher angular wavenumber than the central burning models. As described in Section~\ref{sec:equations1} and in more detail in \citet{Wu2021}, the peak of the wave spectrum in angular wavenumber~$\ell$ occurs for~$\ell \gtrsim \ell_{\rm con} = r/\min(H,\Delta r)$, which is evaluated at the top of the burning shell. Figure~\ref{fig:10v11lcon} shows how the decrease in scale height~$H$ with radius causes the value of~$\ell_{\rm con}$ to increase rapidly as a function of radius in both a centrally-burning ($11\, M_{\odot}$) and off-center burning ($10\, M_{\odot}$) model. Since the off-center burning shell extends to larger radii than a core burning region, waves excited due to Ne burning in a~$10\, M_{\odot}$ model peak at higher~$\ell$. For the snapshot shown,~$\ell_{\rm con} \sim 3~$ for the~$10\, M_{\odot}$ model so its wave spectrum typically peaks at around~$\ell \sim 4$, while~$\ell_{\rm con} \sim 1.5$ for the~$11\, M_{\odot}$ model so its wave spectrum typically peaks around~$\ell \sim 2\text{--}3$. 

The bottom left panel of Figure~\ref{fig:10fesc} demonstrates that the escape fraction is many orders of magnitude smaller for high-$\ell$ waves, shown here for a~$10\, M_{\odot}$ supergiant model but also generally true in all our models. Hence, when the wave spectrum peaks at higher~$\ell$, more of the initial wave power~$L_{\rm wave}$ is distributed to waves that can only escape with a tiny fraction of their original power. The ultimate heating rate~$L_{\rm heat,\ell}$ is dominated by low~$\ell$ waves which only carry~$\dot{E}_{\ell} \lesssim 0.1\, L_{\rm wave}$ (top left panel of Figure~\ref{fig:10fesc}). Since off-center burning excites a wave spectrum that peaks at higher~$\ell$, wave heating is thus strongly reduced.

For the most part, convective excitation of waves in the stripped star models matches that of the supergiant stars, as expected given their nearly equal He core masses. Nevertheless, subtle structural differences in the stripped star models can modify their wave heating rates. In our stripped star models, Ne ignites at a different temperature~$T_{\rm burn, Ne}$ than in the supergiant models. Though both the~$M_{\rm ZAMS} = 10\, M_{\odot}$ stripped star and supergiant models ignite off-center at similar radii, the stripped star model is a little denser and hotter at the center compared to the supergiant when Ne burning begins. Given slightly higher~$T_{\rm burn, Ne}$, convection in the temperature-sensitive Ne burning shell is more energetic, which raises~$L_{\rm con}$,~$\mathcal{M}_{\rm con}$ and consequently~$L_{\rm wave}$ by a factor of a few in the stripped star. 

More vigorous core convection also produces higher~$\omega$, which affects the escape fraction of the~$M_{\rm ZAMS} = 10\, M_{\odot}$ stripped star. Comparing the left and right panels of Figure~\ref{fig:10fesc}, we see that~$f_{\rm esc,\ell}$ is generally higher during Ne burning for the~$M_{\rm ZAMS} = 10\, M_{\odot}$ stripped star. Figure~\ref{fig:10Mprop} shows propagation diagrams for the~$M_{\rm ZAMS} = 10\, M_{\odot}$ supergiant and stripped star models during Ne burning for typical wave frequencies. 
The stripped star model tends to excite higher~$\omega$ by a factor of two, and as demonstrated in the figure, these waves encounter smaller evanescent zones. Thus waves have a larger transmission coefficient $T_{\rm min}^2$ and a larger escape fraction in the stripped star. The combination of increasing both~$f_{\rm esc,\ell}$ and~$L_{\rm wave}$ accordingly raises the wave energy deposited in the envelope by an order of magnitude in the~$M_{\rm ZAMS} = 10\, M_{\odot}$ stripped star (see Figure~\ref{fig:10to13surf}) relative to the H-rich model. For other masses, the difference in wave heating rates between H-rich and H-poor models is much smaller and stems mainly from slight differences in the core structure that affect $T_{\rm burn, Ne}$ and consequently $L_{\rm wave}$ as discussed above.


As discussed in \citet{Wu2021}, our updated implementation of wave physics tends to reduce our wave heating rates by an order of magnitude compared to earlier results such as \citet{fuller2017} and \citet{fuller2018}. Prior work did not model non-linear wave dissipation and assumed that wave power was mainly excited in~$\ell=1$ waves. In particular, the wave heating history of our models is now dominated by a brief burst of very high~$L_{\rm heat,\ell}$ waves excited during Ne burning. We do not produce high wave power sustained for longer periods, such as that during core O burning sustained for~$\sim \! 1$ yr in \citet{fuller2017} and \citet{fuller2018}. 

\subsection{Wave Dissipation}

Both the lower energy scale and sudden, short-duration nature of wave heating influence the hydrodynamical response of our models' interior structures. In the left column of Figure~\ref{fig:11Ebind}, the density and exterior binding energy of the~$11\, M_{\odot}$ supergiant star are plotted as a function of mass coordinate for a model during Ne burning where the instantaneous wave power is~$\sim \! 10^{9}\, L_{\odot}$. The density and effective~$M_{\rm damp}$ (middle panel) drop at the core-envelope transition in the star, so that most of the waves damp their energy at the base of the supergiant envelope at around~$M \sim 3.4 M_{\odot}$. Wave energy is therefore deposited where the binding energy of overlying material is still high,~$E_{\rm bind} \sim 10^{48}\, \mathrm{erg} > E_{\rm waves}$ (see Figure~\ref{fig:10to13surf}, top panel for~$E_{\rm waves}$). The density profile of the supergiant star experiences only minuscule changes due to wave heating, and the deposited wave energy is unable to unbind the overlying envelope mass. Waves in the other supergiant models, which carry much less energy into the envelope than the~$11\, M_{\odot}$ model, likewise deposit their energy without unbinding any material or greatly affecting the density profile of the star.

Our stripped star models lack an overlying hydrogen envelope, so the binding energy is smaller at the edge of the core, as shown in the right panel of Figure~\ref{fig:11Ebind}. Waves again damp most of their energy where the damping mass drops outside the core, which for a stripped star is very close to the surface with only~$M_{\rm ext} \sim 10^{-2}\, M_{\odot}$ of overlying mass. 
For the~$M_{\rm ZAMS} = 11\, M_{\odot}$ model shown in Figure~\ref{fig:11Ebind}, waves with a typical heating rate of $L_{\rm heat, \ell} \gtrsim 10^9\, L_{\odot}$ still damp out where~$E_{\rm bind} \sim 10^{48}$ erg~$> E_{\rm waves}$ since~$M_{\rm damp}$ is smaller for such large $L_{\rm heat, \ell} = L_{\rm ac}$ (Equation \ref{eq:Mdampshock}). Thus the waves excited by Ne burning are not able to unbind the~$M_{\rm ext} \sim 10^{-3} \text{--} 10^{-2}\, M_{\odot}$ of exterior material. Unlike the supergiant models, however, the envelope of this stripped star model inflates above the heating region, flattening the density profile. 
In the other stripped star models, where the wave energy deposited is at least a few times lower (Figure~\ref{fig:10to13surf}), smaller $L_{\rm heat,\ell}$ leads to larger $M_{\rm damp}$, so waves damp further out where~$E_{\rm bind} \sim 10^{47}$ erg. Again,~$E_{\rm wave} \lesssim E_{\rm bind}$ where the waves damp, so waves are also not able to unbind the overlying envelope in the other stripped star models.

The bottom panels of Figure~\ref{fig:11Ebind} show the following timescales for the~$M_{\rm ZAMS} = 11\, M_{\odot}$ supergiant (left) and stripped star (right): the local wave heating timescale,
\begin{equation}
    t_{\rm heat} = \frac{c_s^2}{\epsilon_{\rm wave}}
\end{equation}
where~$\epsilon_{\rm wave}$ is defined in Equation~\ref{eq:dLdm};
the thermal cooling timescale,
\begin{equation}
    t_{\rm therm} = \frac{4\pi\rho r^2 H c_s^2}{L}
\end{equation}
where~$L$ is the local luminosity;
and the local dynamical timescale,
\begin{equation}
    t_{\rm dyn} = \frac{H}{c_s}.
\end{equation}
Both the supergiant and stripped star models lie in the moderate heating regime where~$t_{\rm dyn} < t_{\rm heat} < t_{\rm therm}$. Wave heat is not transported outward thermally, but the local pressure will gradually increase and the star will expand quasi-hydrostatically in response. 
This result differs from \citet{fuller2017}, which also treated wave heating in a supergiant model. \citet{fuller2017} found~$t_{\rm heat}$ could become shorter than both~$t_{\rm therm}$ and~$t_{\rm dyn}$ during Ne burning, so that wave heating entered a dynamical regime and launched a pressure wave propagating at the sound speed. The difference arises because \citet{fuller2017} do not include the weak shock dissipation introduced in \citet{fuller2018}. For our supergiant models with weak shock dissipation, this reduces the effective~$M_{\rm damp}$ of our models compared to \citet{fuller2017} and causes the waves to damp at smaller mass coordinates with smaller values of $t_{\rm dyn}$. As a result, the supergiant models now lie in the moderate heating regime. Our stripped star models remain consistent with \citet{fuller2018}, who also find that~$t_{\rm dyn}<t_{\rm heat} < t_{\rm therm}$ in the wave heating region.

\begin{figure*}
    \centering
    \includegraphics[width=\textwidth]{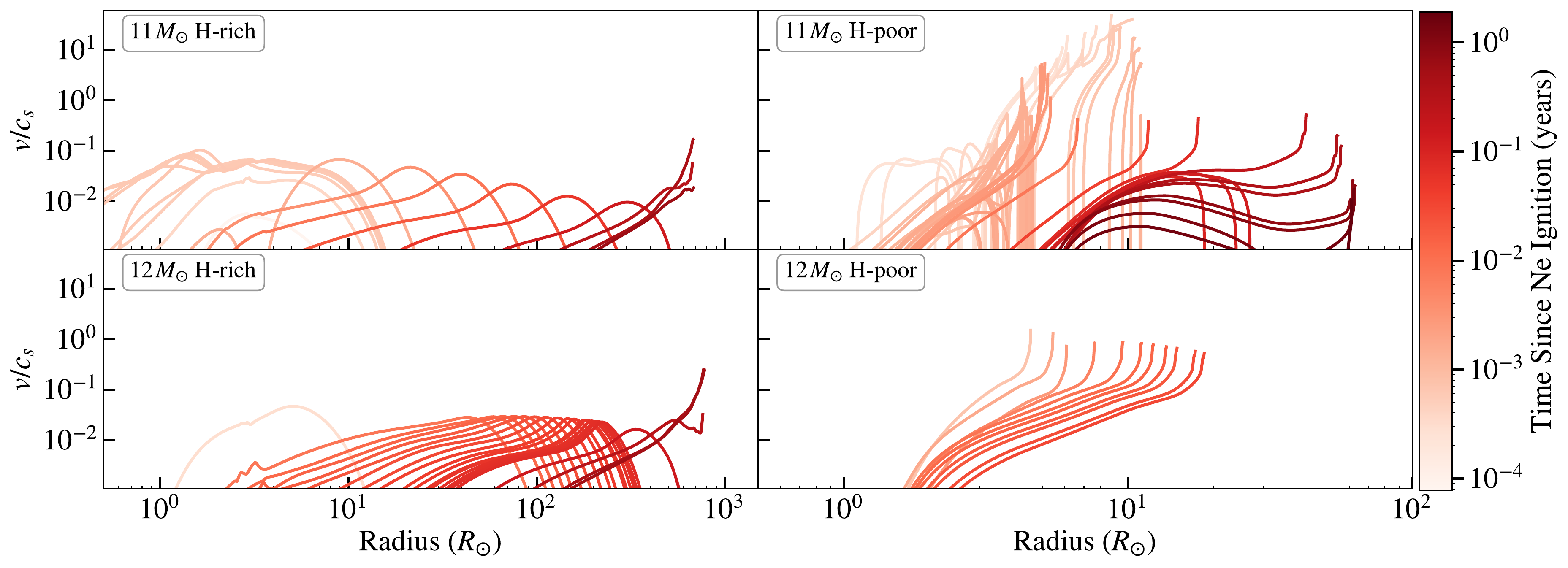}
    \caption{Velocity pulse propagation due to first wave heating phase from Ne ignition in the following models, clockwise from top left:~$M_{\rm ZAMS}=11\, M_{\odot}$ supergiant model,~$M_{\rm ZAMS}=11\, M_{\odot}$ stripped star model, ~$M_{\rm ZAMS}=12\, M_{\odot}$ stripped star model, $M_{\rm ZAMS}=12\, M_{\odot}$ supergiant model. Shading corresponds to the time since Ne ignition. 
    }
    \label{fig:vgrid}
\end{figure*}

\begin{figure*}
    \includegraphics[width=0.515\textwidth]{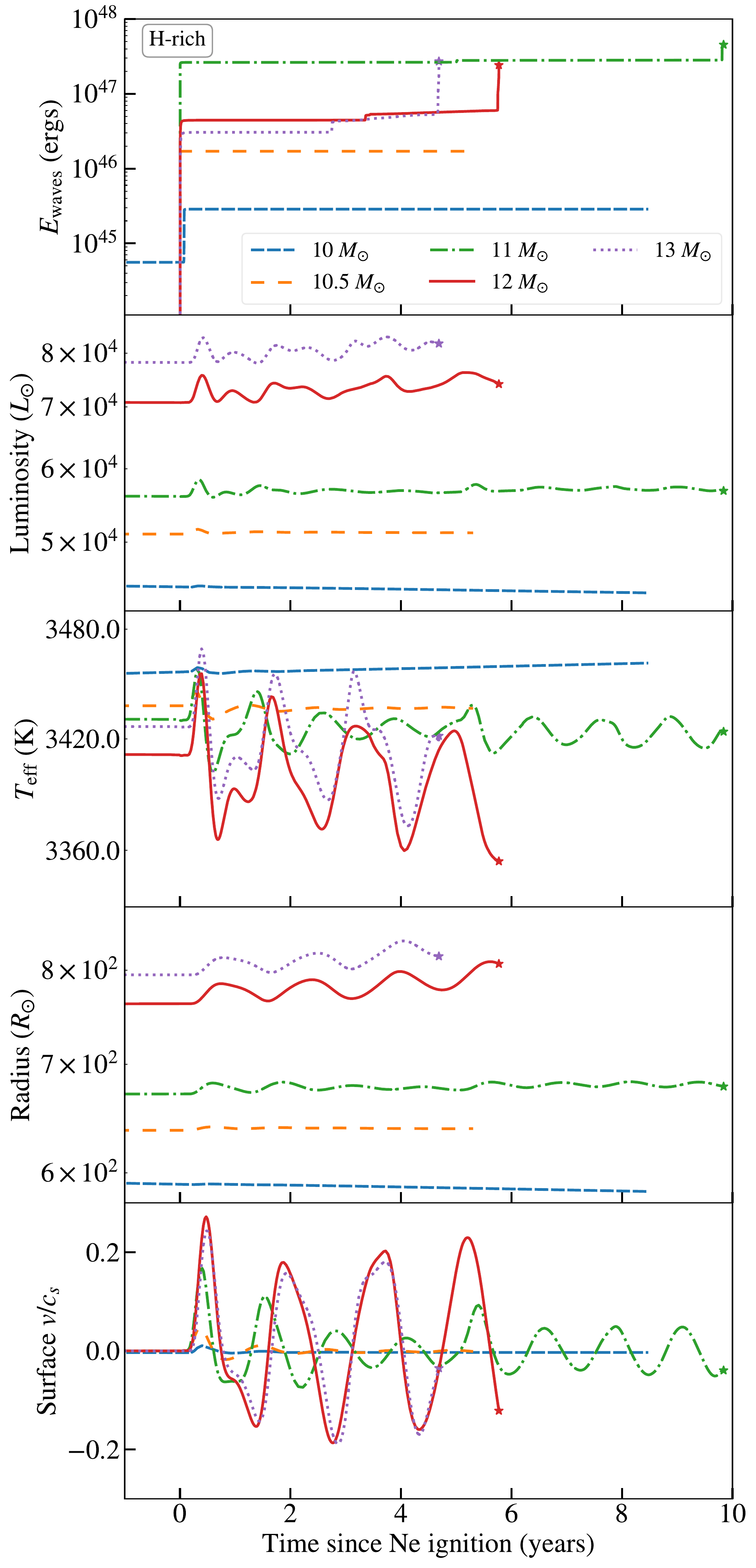}
    \includegraphics[width=0.485\textwidth]{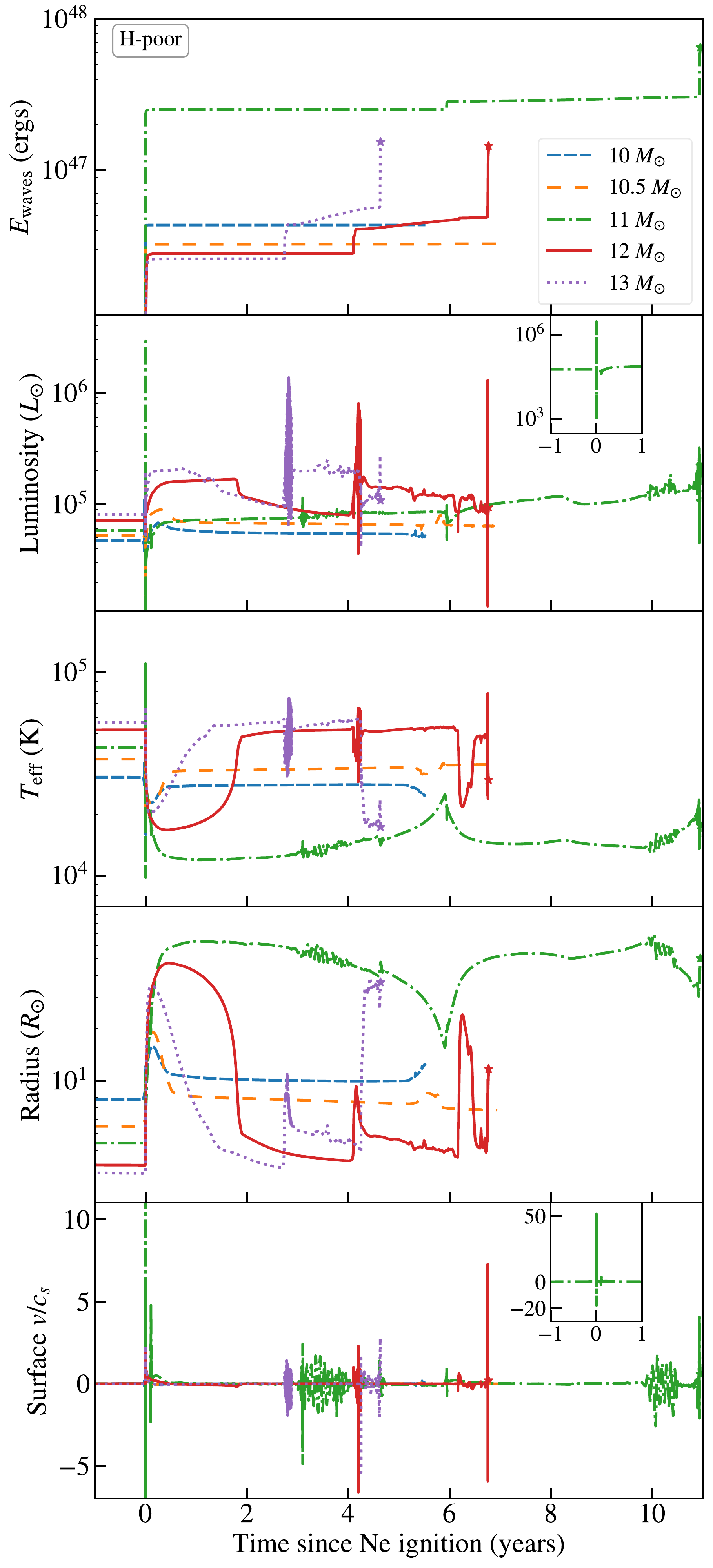}
    \caption{Left: Integrated wave energy deposited, surface luminosity, effective temperature, star radius, and surface velocity as a function of time since Ne ignition for the supergiant models shown in the legend. Right: Same as left panel for the stripped star models with initial masses as shown in the legend. For models that reach core collapse, indicated by the star symbol at the end of each curve, the evolution is shown until 3 days before core collapse. 
    }
    \label{fig:10to13surf}
\end{figure*}

\subsection{Pre-supernova evolution}
\label{sec:results2}

Once wave heat is deposited in our supergiant models, a small fraction of the energy is put into the kinetic energy of a velocity pulse (i.e., a pressure wave) that travels across the envelope of the star.
The left column of Figure~\ref{fig:vgrid} shows the propagation of the velocity pulse due to wave heating from Ne burning in the~$11\, M_{\odot}$ and~$12\, M_{\odot}$ supergiant models. The pulse does not exceed~$v/c_s \sim 0.1$ in either model until it steepens at the surface of the supergiants to~$v/c_s \sim 0.2$. The stars otherwise expand quasi-hydrostatically in response to this velocity pulse. As the surface velocities are quite small, with~$v < v_{\rm esc}$ in all the supergiant models, the surfaces of these stars expand slightly but remain bound.

Note that in our supergiant models, although more wave heat is deposited in the envelope of the~$11\ M_{\odot}$ model and its pressure pulse initially has more kinetic energy, peaking at~$v/c_s \sim 0.1$, the kinetic energy in the pulse decreases greatly as it propagates so that the surface velocity of the~$11\, M_{\odot}$ model is smaller than that of the~$12\, M_{\odot}$ model. In contrast, the pulse in the~$12\, M_{\odot}$ supergiant model begins with a lower peak value because a few times less wave energy was deposited, but it only experiences a very shallow decline before it steepens at the surface.
We attribute this counter-intuitive result to the different profiles of~$t_{\rm therm}$ in the models. As seen in the bottom panel of Figure~\ref{fig:11Ebind},~$t_{\rm therm}$ decreases by a few orders of magnitude at~$m \sim 3.5\, M_{\odot}$, just above the wave heating region at the base of the convective envelope. This dip occurs where the convective flux increases in response to the extra deposited energy (see Section~\ref{sec:TDC} for a discussion of this phenomenon). In the~$11\, M_{\odot}$ model, the small~$t_{\rm therm}$ in this location may allow the convective flux to carry away some energy from the pulse, causing the velocity to decline as in Figure~\ref{fig:vgrid}. The value of~$t_{\rm therm}$ drops less steeply in the~$12\, M_{\odot}$ model, so the local luminosity is not as efficient in carrying energy away from the pressure pulse. As the other models all deposit less wave energy, their velocity pulses emulate the~$12\, M_{\odot}$ model in this respect.

In the~$M_{\rm ZAMS} = 11\, M_{\odot}$ stripped star model shown in the top right of Figure~\ref{fig:vgrid}, the heating generates a rapid expansion 
with Mach number~$\mathcal{M} \sim 50$ near the surface of the star. The response of convection to wave-heating as described above for the supergiants does not apply for the stripped stars, since they lack a convective hydrogen envelope. In the~$M_{\rm ZAMS} = 12\, M_{\odot}$ model (shown in the bottom right of Figure~\ref{fig:vgrid}), less wave heat is deposited and the velocity pulse accelerates to lower values of~$\mathcal{M} \sim 2$. This is representative of the other stripped star models, which deposit comparable amounts of wave energy. Each pressure wave breakout corresponds to spikes in luminosity and effective temperature and initiates rapid envelope expansion of the stripped stars. However, only the expansion driven by the most energetic~$M_{\rm ZAMS} = 11\, M_{\odot}$ model directly after Ne ignition accelerates any stellar material to~$v_{\rm esc} \sim 400$ km/s, and only a very small amount of mass ($M\sim 10^{-6} M_{\odot}$) achieves that velocity. The other stripped star models expand without ejecting any mass.

Figure~\ref{fig:10to13surf} shows the evolution of the surface properties of our models, in parallel with the integrated wave energy deposition as a function of time since Ne ignition. In the supergiant models, the surface luminosity~$L$, effective temperature~$T_{\rm eff}$, radius~$R$, and surface velocity oscillate from Ne ignition onward. Wave heating from Ne ignition initiates these oscillations, which persist until core collapse for the~$11\text{--}13\, M_{\odot}$ models. The amplitude of these fluctuations is small, ranging from almost no variation in the~$10\, M_{\odot}$ model where $E_{\rm wave} \sim 10^{45}$ erg to oscillation amplitudes of a few percent in the~$10.5\text{--}13\, M_{\odot}$ models with $10\text{--}100$ times more wave heating. 

Though our stripped star models mostly deposit comparable amounts of energy in the envelope,
the surface properties of these models behave extremely differently from their supergiant counterparts. The initial pressure pulse breakout originating from wave heating due to Ne ignition generates upward spikes in~$L$ by a factor of a few for most models, although the luminosity of the most energetic~$M_{\rm ZAMS} = 11\, M_{\odot}$ model temporarily jumps by over an order of magnitude. The luminosity in each model also tends to spike downward briefly before becoming brighter than before wave heating began.
Overall, the stripped star models exhibit significant photospheric cooling and expansion, where~$T_{\rm eff}$ drops by up to a factor of a few and the radius of the star rapidly increases by up to an order of magnitude. The properties of the lowest-mass~$M_{\rm ZAMS} = 10\text{--}10.5\, M_{\odot}$ models change less dramatically; for instance, these models expand only by a factor of $2\text{--}3$.

After Ne ignition, a second step-like increase in~$E_{\rm waves}$ occurs for the~$M_{\rm ZAMS} = 11\text{--}13\, M_{\odot}$ models at a few years after Ne ignition. This is due to a second surge of wave heating from core Ne burning, since the burning fuel is replenished when some Ne is mixed downward into the core. This tends to initiate another cooling and expansion phase in the stripped stars, so that over the period between Ne ignition and core collapse, the models exhibit two bumps in radius and dips in~$T_{\rm eff}$. In the~$M_{\rm ZAMS} = 12\text{--}13\, M_{\odot}$ models, the first cool, expanded phase lasts~$\sim 1\text{--}2$ years before returning to near pre-wave heating values; meanwhile, the second Ne burning phase causes the stars to expand less dramatically for a brief period of months. However, these phases last for~$\sim \! 5$ years each in the~$M_{\rm ZAMS} = 11\, M_{\odot}$ model, attaining a similar peak radius and $T_{\rm eff}$ each time. 

Note that this second burning phase also causes spikes in~$L$ and~$T_{\rm eff}$ for the~$M_{\rm ZAMS} = 12\text{--}13\, M_{\odot}$ stripped star models at around 5 and 3 years since Ne ignition respectively.
The models oscillate in~$L$ and~$T_{\rm eff}$ because the Ne abundance in the core is fluctuating. This drives oscillations in the burning rate and~$L_{\rm con}$, which in turn causes the wave heating rate and surface properties to fluctuate.
Increasing the resolution of our models did not consistently remove this effect, so it is unclear whether this effect is physical or numerical in nature. 
Different treatments of convective overshoot could affect the core Ne abundance as well. 

Although we predict little hydrodynamically-driven mass loss, the large changes in surface luminosity exhibited by the stripped star models affects the wind mass loss rates of these stars. Using the following wind mass loss rate from \citet{Nugis2000},
\begin{equation}
\label{eq:Mdotwind}
    \begin{split}
   \log \dot{M}_{\rm wind} = & -11 + 1.29\log L  \\
   & + 1.73 \log Y + 0.47 \log Z,
\end{split}
\end{equation}
we calculate the expected mass loss rate from winds in each of the stripped star models due to their elevated luminosity after Ne burning. We find that the increased mass loss from winds is an order of magnitude larger than directly ejected material, but still very small. The wind mass loss rates in our models peak at~$\dot{M} \sim 10^{-5}\, M_{\odot}\text{/yr}$, and the models lose a few$\times 10^{-5}\, M_{\odot}$ between Ne ignition and core collapse.

\section{Discussion}
\label{sec:discussion}

\subsection{Parameter testing}
\label{sec:paramtest}

\begin{figure}
    \centering
    \includegraphics[width=0.49\textwidth]{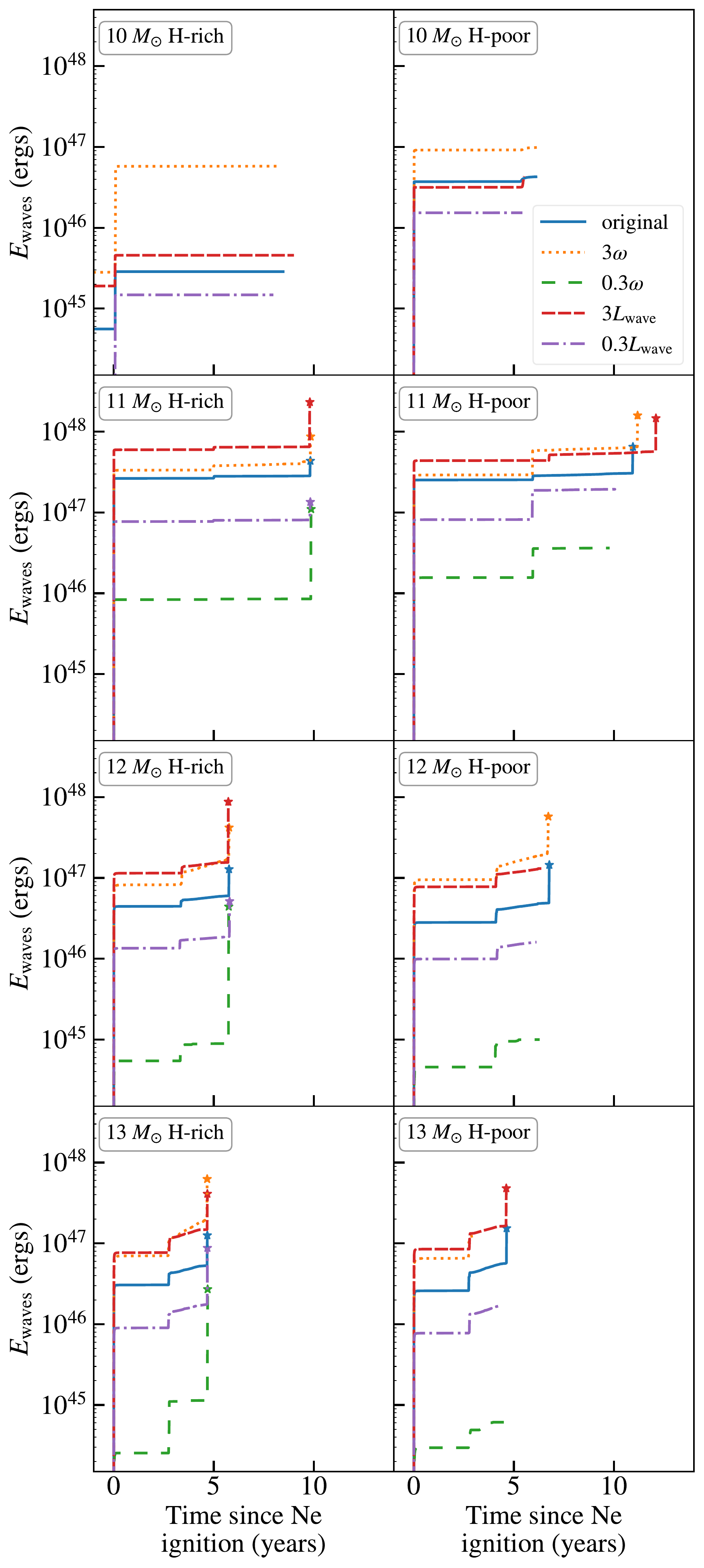}
    \caption{Integrated wave energy deposited as a function of time since neon ignition for supergiant (left column) and stripped star (right column) models with the initial masses shown in each panel. Each panel shows the variation in wave energy deposition for each model as either the wave frequency (Equation~\ref{eq:omega}) or the power put into waves (Equation~\ref{eq:lwave}) is increased and decreased by a factor of 3. As in Figure \ref{fig:10to13surf}, models that reach core collapse are indicated by the star symbol at the end of each curve. }
    \label{fig:Ewavesparam}
\end{figure}

To investigate how our results may vary if our assumptions about wave power and frequency do not accurately represent the true values, we ran simulations with different values for~$L_{\rm wave}$ (Equation~\ref{eq:lwave}) and~$\omega$ (Equation~\ref{eq:omega}). For each model in Figure~\ref{fig:10to13surf}, we repeated the methods of Section~\ref{sec:methods}, each time making one of the four following changes:
\begin{enumerate}[noitemsep]
    \item~$\omega = 3\omega_{\rm con}$
    \item~$\omega = 0.3 \omega_{\rm con}$
    \item~$L_{\rm wave} = 3 \mathcal{M}_{\rm con} L_{\rm con}$
    \item~$L_{\rm wave} = 0.3 \mathcal{M}_{\rm con} L_{\rm con}$.
\end{enumerate}
These variations allow us to test how our results would change if our initial assumptions for the power and frequency of the excited waves either overestimate or underestimate reality. With these simulations along with our original models, we span an order of magnitude of uncertainty in these initial assumptions. Note that we do not change our assumed wave spectrum for these tests, so changes in~$\omega$ only affect the dispersion relation and related quantities outlined in Section~\ref{sec:physics}.

\begin{figure*}
    \centering
    \includegraphics[width=0.51\textwidth]{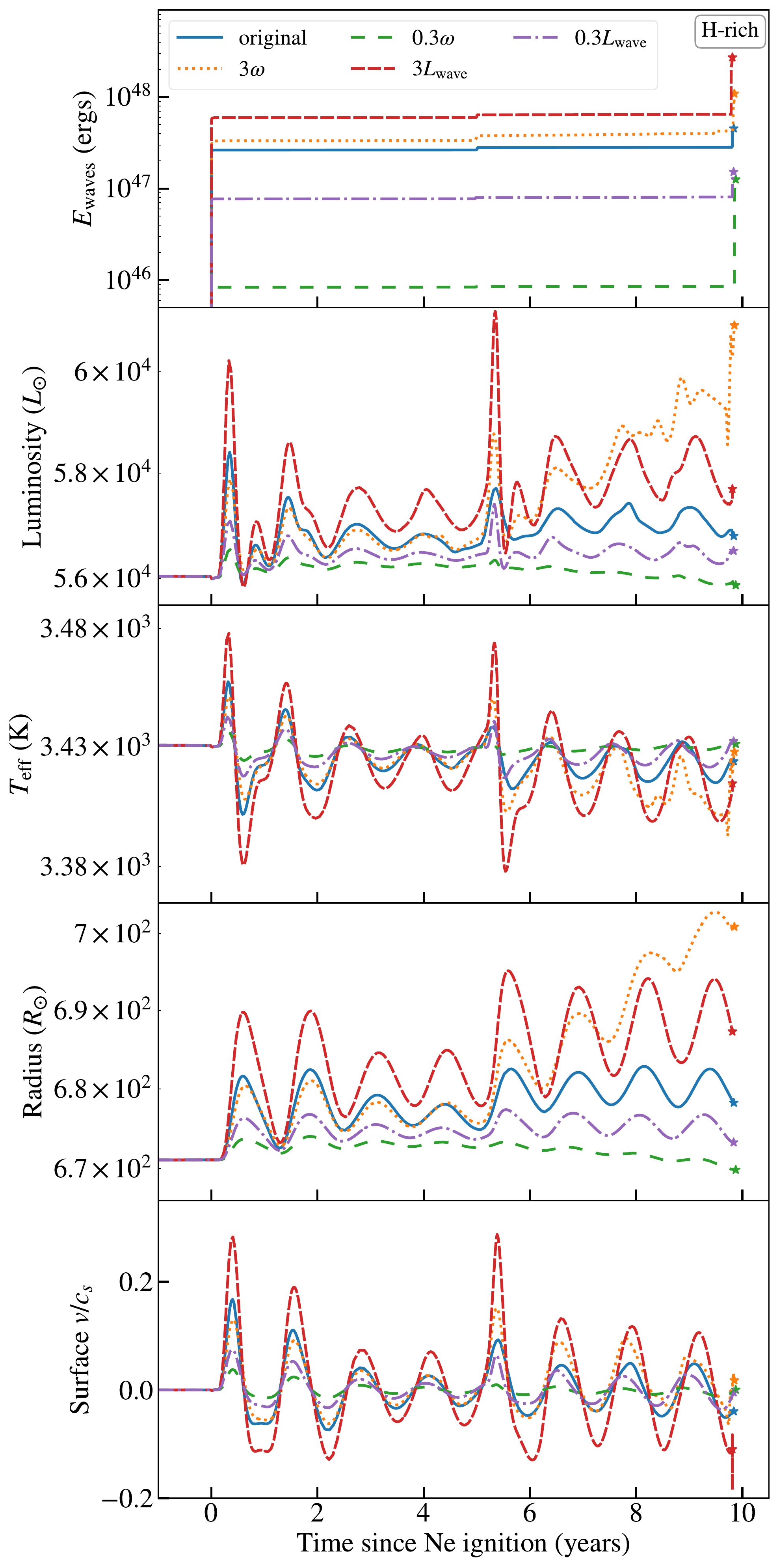}
    \includegraphics[width=0.47\textwidth]{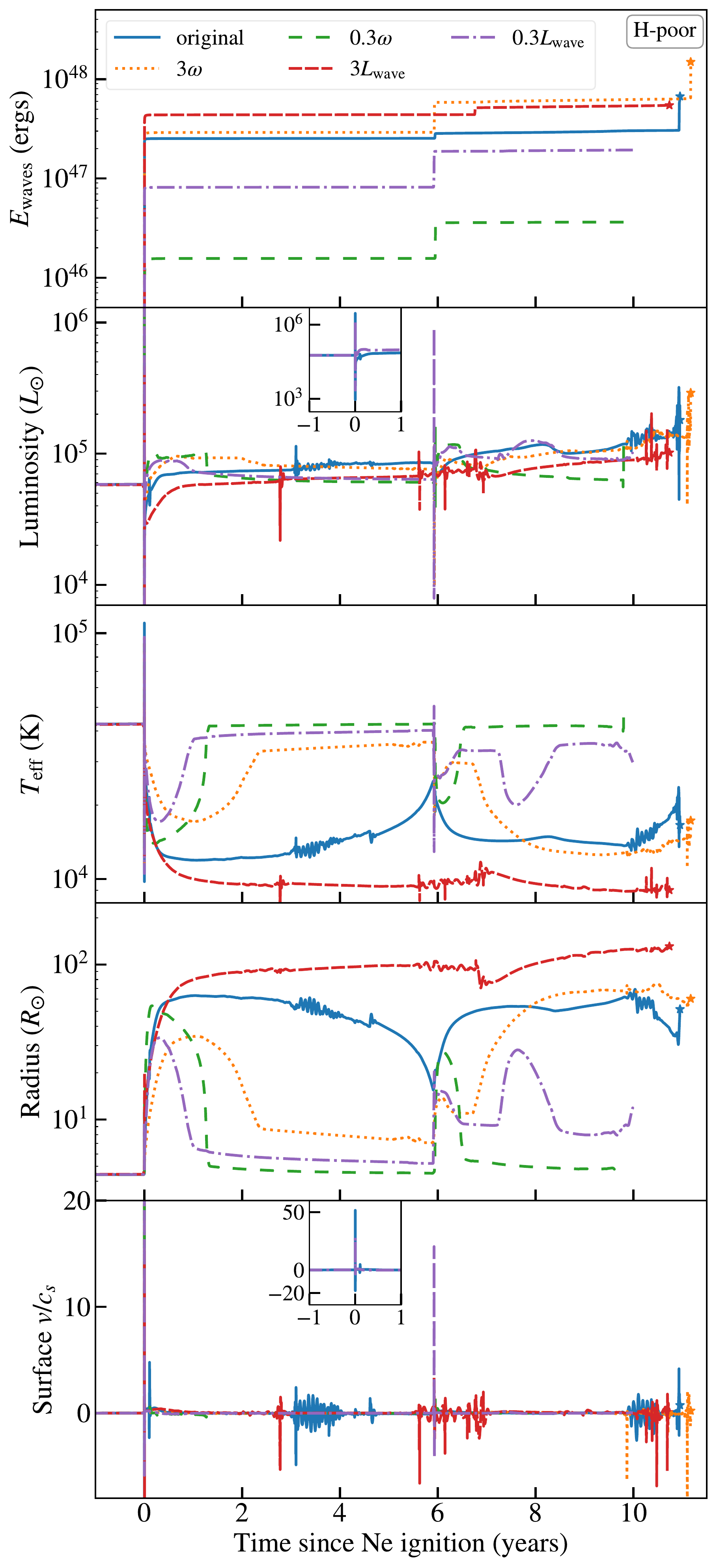}
     \caption{Left panel: Same quantities as in Figure~\ref{fig:10to13surf} are shown for an~$11\, M_{\odot}$ supergiant model. As in Figure~\ref{fig:Ewavesparam}, each panel plots the different evolution of each model as either the wave frequency or power is varied (see Section~\ref{sec:paramtest} for details). Right panel: Same as left for the~$M_{\rm ZAMS}=11\, M_{\odot}$ stripped star model. As in Figure \ref{fig:10to13surf}, models that reach core collapse are indicated by the star symbol at the end of each curve. 
    }
    \label{fig:11surfparam}
\end{figure*}

Figure~\ref{fig:Ewavesparam} shows how the wave energy that escapes to heat the envelope changes as we vary these parameters in each model. We omit~$M_{\rm ZAMS} = 10.5\, M_{\odot}$ as its evolution due to off-center Ne ignition proceeds similarly to~$M_{\rm ZAMS} =10\, M_{\odot}$. For all models, the wave heating history remains the same -- jumps in~$E_{\rm waves}$ occur at the same times, as changing~$L_{\rm wave}$ or~$\omega$ has no bearing on when the model develops convective regions.
As noted in \citet{Wu2021}, wave energy transmission in the mass range~$11\text{--}15\, M_{\odot}$ is not significantly altered by non-linear effects, so for these models the wave energy escape generally increases (decreases) by a factor of~3 as we increase (decrease)~$L_{\rm wave}$ by a factor of~3. However, the~$3 L_{\rm wave}$ variations upon each of the~$M_{\rm ZAMS} = 10\, M_{\odot}$ models and the~$M_{\rm ZAMS} =11\, M_{\odot}$ stripped star model do not show as large of an increase in wave energy; in fact, slightly less wave energy escapes for the~$M_{\rm ZAMS} =10\, M_{\odot}$,~$3 L_{\rm wave}$ variation than the original run. This is due to non-linear effects in these models -- increasing~$L_{\rm wave}$ also increases the wave non-linearity~$|k_r\xi_r|^2$ by the same factor, which for non-linear waves effectively reduces the wave power back to that of the original run.

Changing~$\omega$ affects the escape fraction and wave non-linearity. In the core, higher frequency waves will experience less attenuation due to neutrino and thermal losses in the g-mode cavity, and they have larger escape fractions. 
The non-linearity factor goes as~$|k_r\xi_r|^2 \propto \omega^{-4}$, so higher frequency waves are much less non-linear. Thus, we ultimately find that increasing (decreasing) wave frequency increases (decreases) the wave energy escape, in some models more drastically than others depending on the magnitude of the change in~$f_{\rm esc,\ell}$ and whether non-linear effects are important for those waves.

The variation in surface properties as we vary~$\omega$ and~$L_{\rm wave}$ is shown for an~$11\, M_{\odot}$ supergiant model in the left panel of Figure~\ref{fig:11surfparam}. All the models exhibit oscillations of varying amplitude which persist until core collapse, with larger jumps occurring soon after wave heating episodes. The amplitude of the oscillation tends to be larger for variations with more wave heating such as~$3 L_{\rm wave}$ and~$3 \omega$ and much smaller for less energetic models like~$0.3 L_{\rm wave}$ and~$0.3 \omega$. Moreover, the more energetic runs tend towards larger~$L$ and~$R$ and smaller~$T_{\rm eff}$ than the original run, whereas the surface properties of the less energetic variations deviate less from the quiescent values than the original run.

In contrast, the stripped star with~$M_{\rm ZAMS}=11 M_{\odot}$ exhibits much larger changes in its surface properties due to the variations in $\omega$ and $L_{\rm wave}$ (Figure~\ref{fig:11surfparam}, right panel). The luminosity may vary by a factor of~$\sim$2 in either direction compared to the original run, but more energetic models (larger~$L_{\rm wave}$ or~$\omega$) are generally brighter, with larger photospheric radii and lower temperatures.
The~$3 L_{\rm wave}$ run cools and expands more than the original run right after Ne ignition, remaining larger and cooler than all other runs until core collapse. In the least energetic~$0.3 L_{\rm wave}$ and~$0.3 \omega$ runs, wave heating causes the star to initially cool by a factor of~$\sim 2\text{--}2.5$ and expand to tens of~$R_{\odot}$ before returning to radii and temperatures near that of quiescence after about a year.
The behavior of the~$11 \, M_\odot$ models shown in Figure~\ref{fig:11surfparam} qualitatively represents that of the other masses, though variations in the~$10 \, M_\odot$,~$10.5 \, M_\odot$,~$12 \, M_\odot$, and~$13 \, M_\odot$ models are on somewhat smaller scales since less wave heat is deposited in the envelopes of those models.

\begin{figure}
    \includegraphics[width=\columnwidth]{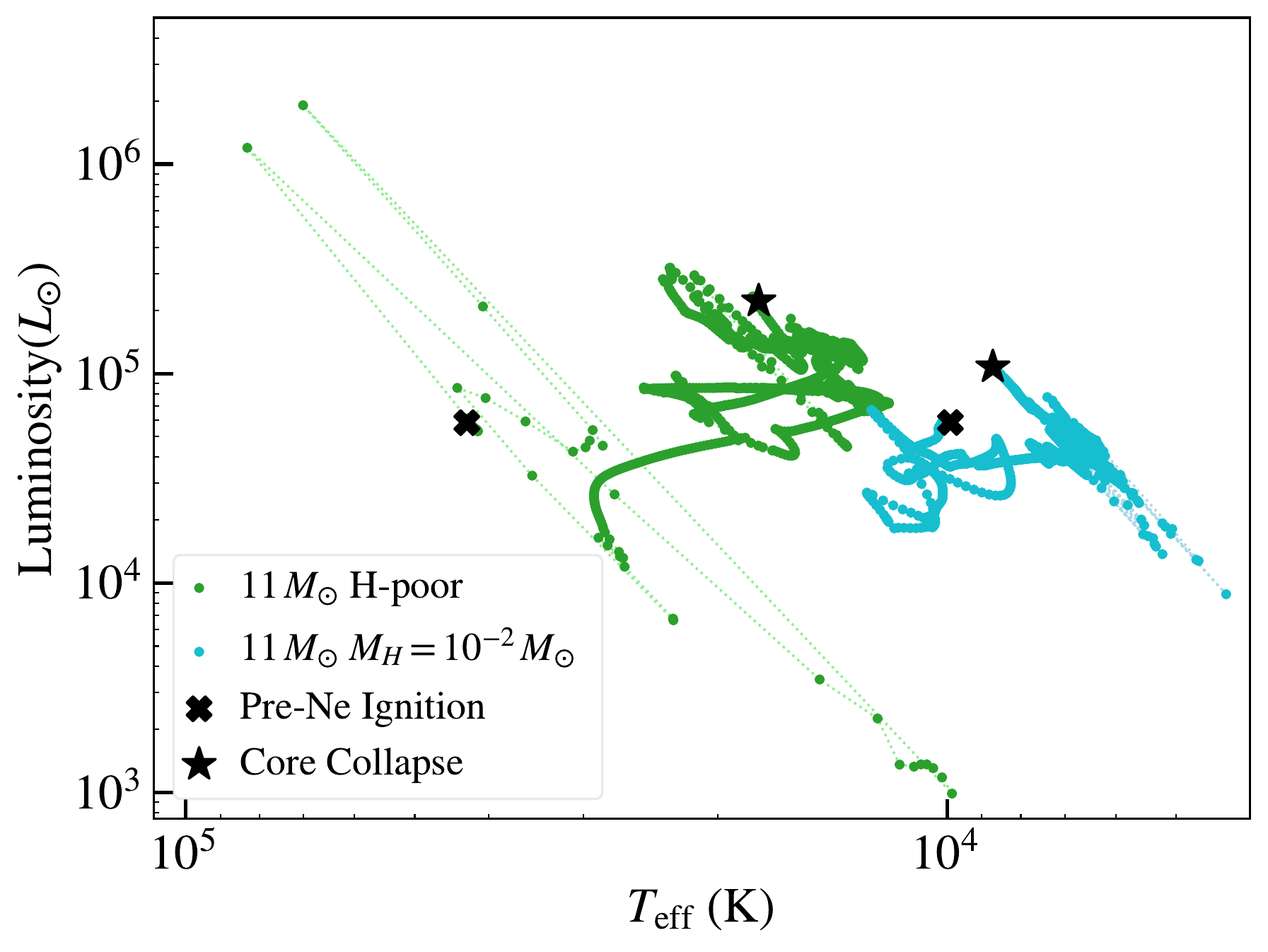}
    \caption{Evolution on the Hertzsprung-Russell diagram for the H-poor~$M_{\rm ZAMS} = 11\, M_{\odot}$ model (green) and a~$M_{\rm ZAMS} =11\, M_{\odot}$ stripped star model with~$M_H = 10^{-2}\, M_{\odot}$ (cyan) from just before Ne ignition (cross) until core collapse (star). Each dot is separated by an interval of 1 hour.
    }
    \label{fig:11MHR}
\end{figure}

\begin{figure}
    \includegraphics[width=\columnwidth]{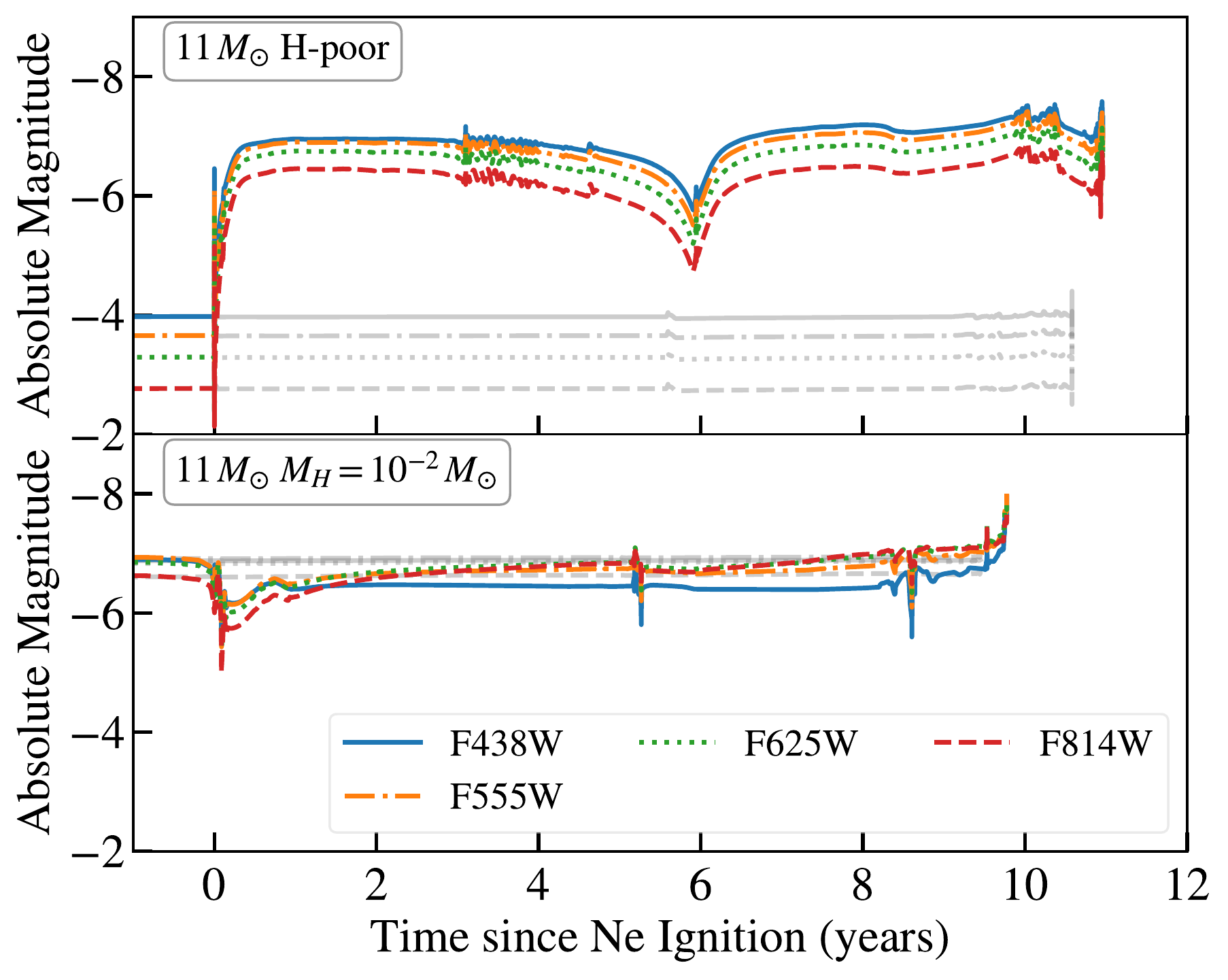}
    \caption{Top: Evolution of absolute magnitude (AB mag) in four HST bandpasses as a function of time since Ne ignition for the H-poor~$M_{\rm ZAMS} = 11\, M_{\odot}$ model. The absolute magnitudes of the same model with no wave heating are plotted in grey, with line styles corresponding to the same line styles of each bandpass in the legend. Bottom: Same for a~$M_{\rm ZAMS} =11\, M_{\odot}$ stripped star model with~$M_H = 10^{-2}\, M_{\odot}$.
    }
    \label{fig:11Mmag}
\end{figure}

\begin{figure}
    \includegraphics[width=\columnwidth]{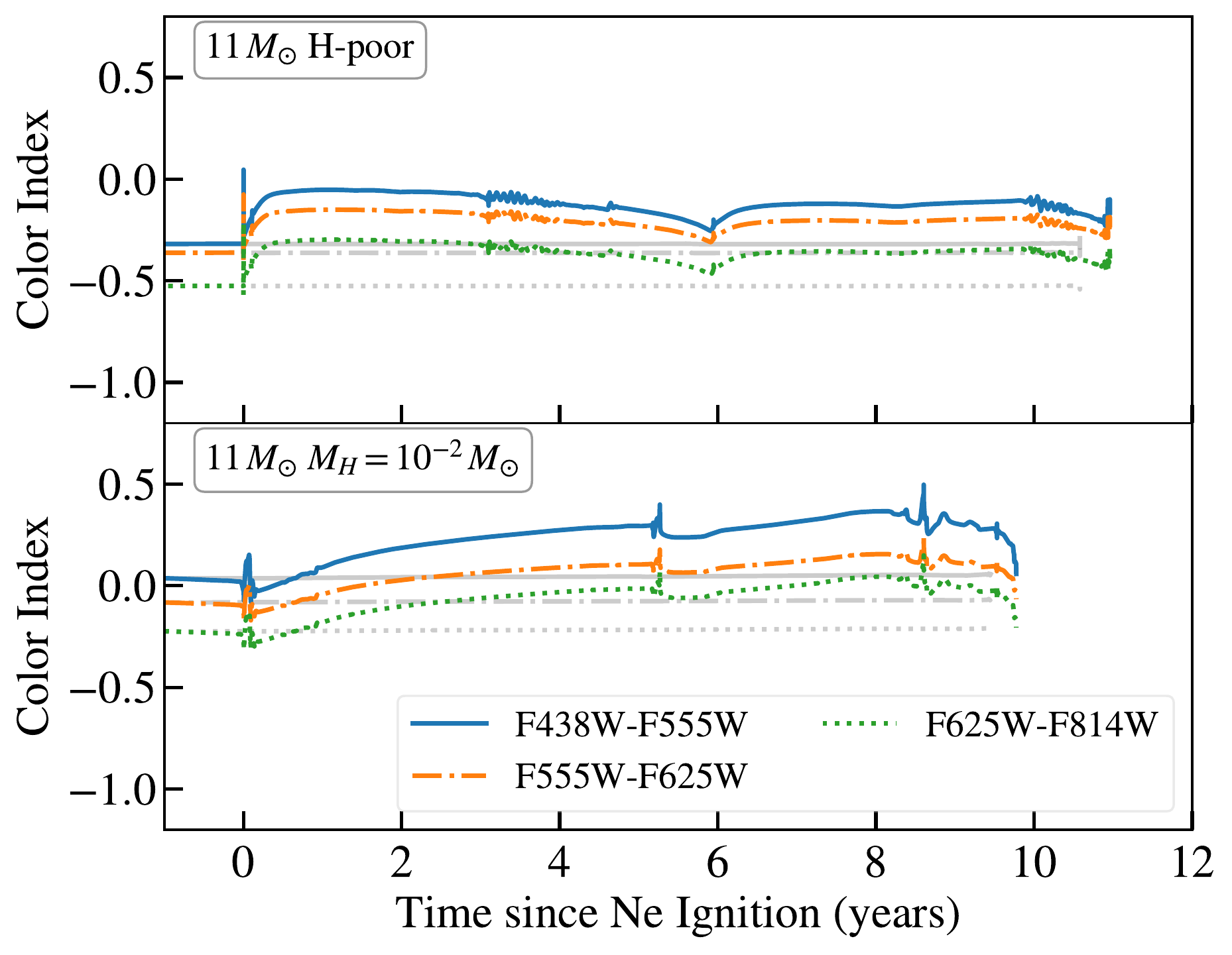}
    \caption{Top: Evolution of several color indices (AB mag) as a function of time since Ne ignition for the H-poor~$M_{\rm ZAMS} = 11\, M_{\odot}$ model. The colors of the same model with no wave heating are plotted in grey, with line styles corresponding to the same line styles of each bandpass in the legend. Bottom: Same for a~$M_{\rm ZAMS} =11\, M_{\odot}$ stripped star model with~$M_H = 10^{-2}\, M_{\odot}$.
    }
    \label{fig:11Mcolor}
\end{figure}

\begin{figure}
    \includegraphics[width=\columnwidth]{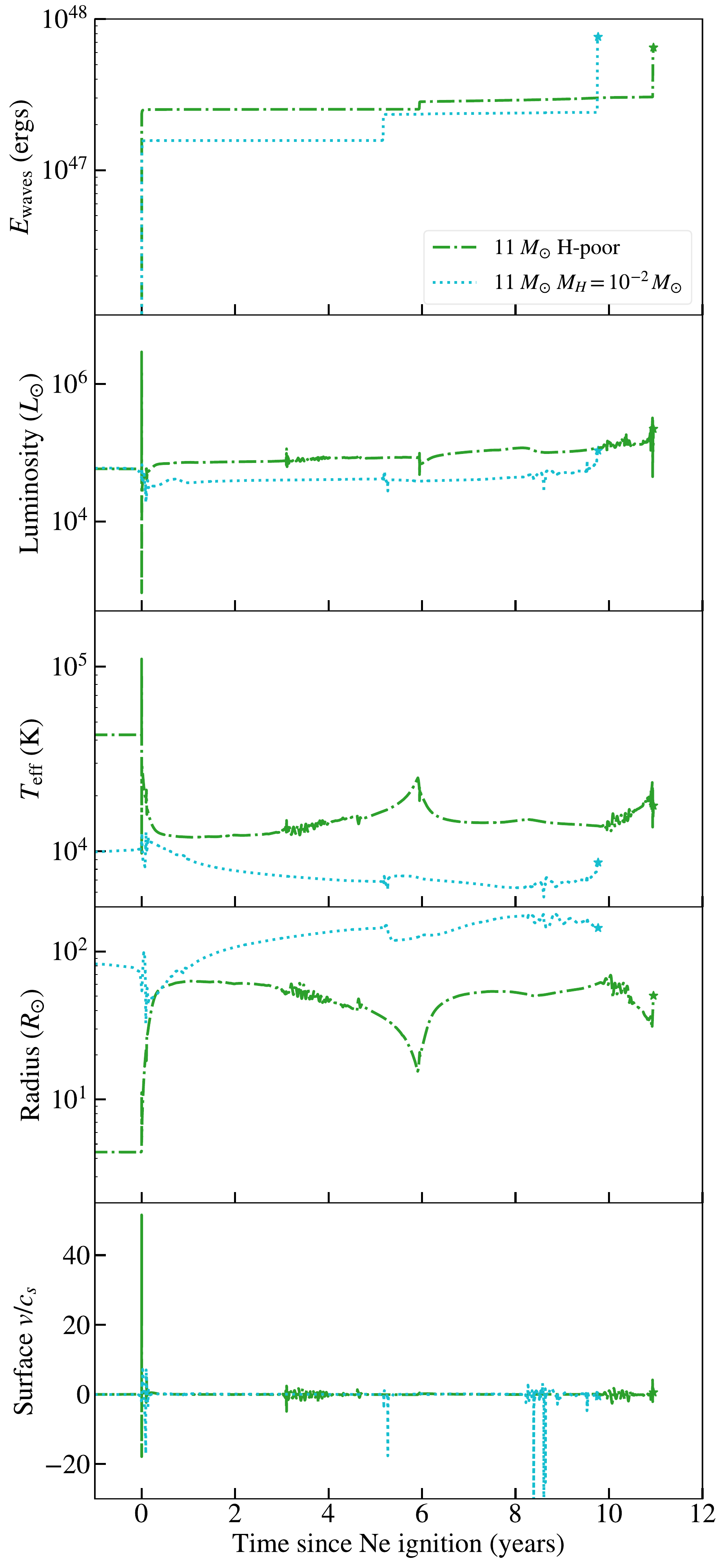}
    \caption{Same quantities as in Figure~\ref{fig:10to13surf} are shown for the H-poor~$M_{\rm ZAMS} =11\, M_{\odot}$ model and a~$M_{\rm ZAMS} =11\, M_{\odot}$ stripped star model with~$M_H = 10^{-2}\, M_{\odot}$. The star symbol at the end of each curve indicates that both models reach core collapse.
    }
    \label{fig:11moreHsurfparam}
\end{figure}

\subsection{Comparison with Progenitor Observations}

Since the changes in the surface properties of our supergiant models are quite small, we do not expect a large observational signature of wave heating in these stars. However, our stripped star models start out as hot, compact He stars, but expand greatly and become cooler for years due to wave heating. Figure \ref{fig:11MHR} shows the evolution of the~$M_{\rm ZAMS} = 11\, M_{\odot}$ stripped star model on the HR diagram from just before Ne ignition until core collapse. A model without wave heat would remain at the point labeled ``Pre-Ne Ignition'' on the green curve in Figure \ref{fig:11MHR} until core collapse. In contrast, the star initially moves to large~$L$ and~$T_{\rm eff}$, then to small~$L$ and~$T_{\rm eff}$, over a few hours when the shock launched by wave heating breaks out. It then spends months moving to a more luminous but cooler state. The star remains in that region of the HR diagram for years, with some continued variability due to fluctuating amounts of wave heat emerging from the core.

The top panel of Figure \ref{fig:11Mmag} shows lightcurves in four \textit{Hubble Space Telescope} WFC3/UVIS bands as a function of time since Ne ignition for this model. Each absolute magnitude was calculated by integrating over a blackbody at~$T_{\rm eff}$ convolved with the transmission function for each bandpass filter and is reported in AB magnitudes. The brightness in all bands jumps once wave heating begins at Ne ignition. Note that the bolometric absolute magnitude will only decrease by~$\sim$1, but the visual band magnitudes decrease by~$\sim$3 magnitudes. This is a consequence of the star cooling from~$\approx$40,000 K to~$\approx$15,000 K, causing a much larger fraction of its flux to be emitted in optical bands rather than the UV.
Like its brightness, the color of our~$M_{\rm ZAMS}=11 \, M_\odot$ H-poor model varies over the last few years of its lifetime. The top panel of Figure \ref{fig:11Mcolor} shows the evolution of some color indices for this star, again in AB mags. Before Ne ignition the star is brightest in UV, but due to wave heating the star reddens (but remains fairly blue) with F438W$-$F555W rising to near zero.

A few potential Type Ib/c supernova progenitors have been observed and are indeed bluer than type II SN progenitors, but in some cases they are cooler or more extended than a typical massive helium star \citep{Cao2013,Xiang2019,Kilpatrick2021}. Among our stripped-star models, which represent type Ib SN progenitors, certain periods of evolution could reproduce these characteristics. For example, \citet{Eldridge2015} find that the progenitor candidate of iPTF13bvn has brightness between $\mathrm{M}_{\rm F435W} = -6.15$ to $-6.67$ mag, $\mathrm{M}_{\rm F555W} = -6.1$ to $-6.49$ mag, and $\mathrm{M}_{\rm F814W} = -5.95$ to $-6.13$ mag. Our~$M_{\rm ZAMS}=11\, M_{\odot}$ H-poor model can match these ranges at around 6 months--5 years after Ne ignition.


The same~$M_{\rm ZAMS}=11\, M_{\odot}$ H-poor model can match the brightness of $\mathrm{M}_{\rm F555W} \gtrsim -5.5$ mag observed by \citet{Kilpatrick2021} for the progenitor of SN2019yvr during most of the 10 years after Ne ignition. However, \cite{Kilpatrick2021} finds the potential progenitor to be much cooler than what our hydrogen-poor models reach, deriving~$T_{\rm eff} = 6800^{+400}_{-200}$ K from models which account for both host and Milky Way extinction. They also report observations of $\mathrm{M}_{\rm F555W}-\mathrm{M}_{\rm F814W} = 1.065\pm0.045$ mag, but this accounts only for Milky Way extinction and not yet host extinction. Nevertheless, based on these values, our ~$M_{\rm ZAMS}=11\, M_{\odot}$ H-poor model would not be able to explain the SN2019yvr progenitor, as it only attains $\mathrm{M}_{\rm F555W}-\mathrm{M}_{\rm F814W} \sim -0.5$ mag.
\citet{Sun2021} find that if the source in \citet{Kilpatrick2021} is a binary system, then the progenitor of SN2019yvr could have~$\log( T_{\rm eff}/K) = 4.03$ and~$R/R_{\odot} = 57$ with likely initial mass~$\sim \! 11\, M_{\odot}$. Our~$M_{\rm ZAMS}=11\, M_{\odot}$ stripped star model matches these observational constraints fairly well for the last 10 years of its life. Yet in both cases, the small amount of progenitor variability over~$\sim$100 days of HST observations \citep{Kilpatrick2021} may disfavor the wave-heated progenitor interpretation.

A stripped star model with a very small amount of hydrogen remaining in the envelope might explain cooler progenitors, as even a hundredth of a solar mass of hydrogen in the envelope causes the star to expand more and reach lower temperatures \citep{Laplace2020}. In our fiducial stripped star models, we strip the star of its envelope so that a negligible amount of hydrogen remains in the envelope ($M_H \lesssim 10^{-3}\, M_{\odot}$) from carbon burning onward. In Figure \ref{fig:11moreHsurfparam}, we compare our fiducial~$M_{\rm ZAMS} = 11\, M_{\odot}$ H-poor model with an~$M_{\rm ZAMS} = 11\, M_{\odot}$ stripped star model with~$M_H = 10^{-2}\, M_{\odot}$ of hydrogen in the envelope from carbon burning onward, evolving the hydrodynamic response to wave heating as in the rest of our models. Even before wave heating begins, the~$M_{\rm ZAMS} = 11\, M_{\odot}$ stripped star with~$M_H = 10^{-2}\, M_{\odot}$ expands to~$\sim 80\, R_{\odot}$ just before Ne burning. However, its luminosity just before Ne burning matches that of the H-poor model, and accordingly is a factor of~$4$ times cooler before Ne burning. This star would also remain at the point labeled ``Pre-Ne Ignition'' on the cyan curve in Figure \ref{fig:11MHR} until core collapse in the absence of wave heating.

The wave heating history of the two models, shown in the top panel of Figure \ref{fig:11moreHsurfparam}, is similar, but the model with~$M_H = 10^{-2}\, M_{\odot}$ cools and expands in response to wave heating from Ne burning so that it is slightly less luminous than the H-poor model until core collapse (Figure \ref{fig:11MHR}). 
Since the~$M_H = 10^{-2}\, M_{\odot}$ model expands to more than~$100 \, R_\odot$, its photospheric cools to the range~$T_{\rm eff} = 6800^{+400}_{-200}$~K derived by \citet{Kilpatrick2021} for the progenitor of SN2019yvr. Consequently, the model with~$M_H = 10^{-2}\, M_{\odot}$ is redder than the fiducial model (Figures \ref{fig:11Mmag} and \ref{fig:11Mcolor}), though it can also explain the color of iPTF13bvn's progenitor estimated in \citet{Eldridge2015}. The model with more hydrogen exemplifies the varied evolution that can ensue depending on the residual hydrogen content of the stripped progenitor, which depends sensitively on the binary and wind mass loss that created the SN progenitor.

\subsection{Related Mass Loss Mechanisms}
\label{sec:massloss}


An interesting aspect of the stripped star models is their expansion from a few~$R_{\odot}$ to tens of~$R_{\odot}$.
As stripped stars are likely to be in a binary system, this substantial, fast expansion may instigate binary mass transfer. The resulting rapid mass transfer could drive more mass loss from the system, or it could lead to unstable mass transfer that results in a common envelope event or stellar merger (see also \citealt{Mcley2014}). Since the mass loss is triggered in the final years before core-collapse, the ejected mass would still be near the progenitor at the time of core-collapse. Hence, this wave-induced binary interaction is a potential channel for generating larger CSM masses that should be explored in future work.

In our~$10$ and~$10.5\, M_{\odot}$ models, we investigated only the initial Ne ignition and the progress of the ensuing off-center flame for~$6\text{--}8$ years after Ne ignition, which encapsulated the majority of wave heating due to Ne burning. Yet this mass range may exhibit interesting behavior beyond this period of its evolution, as degenerate silicon ignition is highly energetic. On the one hand, this silicon burning could excite waves with very high wave power which have greater potential to eject mass. On the other hand, \citet{Woosley2015} have found that silicon deflagration can occur in stars of this mass range, in which the silicon flash is violent enough to drive a shock and eject mass. Though it is computationally expensive to simulate late-stage nuclear burning this mass range, the period of Si burning should be studied in more detail as a possible mechanism for driving mass loss in these stars. 

\subsection{Time Dependent Convection}
\label{sec:TDC}

In our supergiant models, the internal luminosity can suddenly spike by a few orders of magnitude at the base of the convective envelope due to wave heating in this region. This manifests in the bottom panel of Figure~\ref{fig:11Ebind}, where~$t_{\rm therm}$ dips at a mass coordinate of~$3.5\, M_{\odot}$ -- which is due to an associated jump in the luminosity of the model at that location.
This is caused by a sudden increase in the convective luminosity on a time scale comparable to the convective turnover time scale, which may not be physical,
and calls for a time-dependent treatment of convection.
In the results presented in this work, we have tried to partially mitigate unphysically large convective accelerations by using the MESA option \texttt{mlt\_accel\_g\_theta = 1} in our supergiant models, which limits convective acceleration to the local gravitational acceleration. However, a more rigorous treatment of time-dependent convection (TDC) could further limit convective acceleration. This will affect whether wave heat is transported outward by convection, or whether it can be used to hydrodynamically drive a pressure pulse through the star.

To try to understand the range of possibilities, we investigated the efficacy of more sophisticated schemes to limit convective acceleration and deceleration. Following  \citet{Renzo2020}, we tried limiting convective acceleration based on the methods of \citet{Wood1974}. This method failed to create a physical model as spurious drops in luminosity at the surface of the convective envelope occurred. We also attempted the second treatment of TDC in \citet{Renzo2020}, in which they use MESA v11123 to solve for convective velocity using their Equation 2 instead of using MLT. Unfortunately, a model using this method was unable to evolve through the extremely energetic degenerate Ne ignition.

Finally, we tested the response of a supergiant star's convective envelope to added heat at the base of the envelope using an implementation of TDC that is available in a recent development version of MESA.
A model using this TDC treatment exhibits almost identical behavior in the convective envelope to that of the models presented here, but with more numerical difficulties that make it difficult to apply to many models or to run them until core-collapse. Consequently, we choose to present our models that limit convective acceleration by \texttt{mlt\_accel\_g\_theta = 1} and successfully run to core collapse, with the understanding that the convective behavior appears similar with more sophisticated convective acceleration schemes. However, from preliminary tests, the surface properties of the TDC model differed slightly from the results presented here. Moreover, it is possible that TDC could affect the progression of core convection at Ne ignition, which has implications for the wave heating results. We therefore emphasize the need for further work on how TDC affects the evolution of massive stars.

\section{Conclusion}

We have modeled the effect of wave heating in red supergiant and stripped-star SN progenitors with ZAMS masses between~$10\text{--}13\, M_{\odot}$ using one-dimensional hydrodynamical simulations in MESA. In our models, we implement the improved wave heating physics of \citet{Wu2021} as we evolve to core collapse for~$M_{\rm ZAMS} > 11\, M_{\odot}$, replicating the wave energy results of that work. We additionally study models with~$M_{\rm ZAMS} \lesssim 10.5\, M_{\odot}$ which experience off-center ignition of Ne burning and are evolved to a few years after Ne ignition. These transmit much less wave heat than the centrally burning models because the wave power is carried primarily by~$\ell \gtrsim 3$ waves that are mostly damped in the core.

The most energetic model in this study is the~$M_{\rm ZAMS} = 11\, M_{\odot}$ model, which transmits~$\sim \! 10^{47}~$erg of wave energy to its envelope around~$10$ years before core collapse in a burst of vigorous wave heating during central Ne burning. In our other models, waves are able to carry much less energy to the envelope. The wave energy in all of our supergiant and stripped star models is deposited just outside the core where the binding energy of overlying material is still high, and no mass loss is hydrodynamically driven by wave heating in any model. As a result, we do not predict that any of our models can produce significant CSM masses through wave heating alone.

As wave heat is deposited at the base of the hydrogen envelope in our supergiant models, their photospheric properties respond to wave heating with only small oscillations in the final years before core-collapse. In our stripped star models, where wave heat is deposited much closer to the surface of the star, wave heating causes significant photospheric cooling and radial expansion in which~$T_{\rm eff}$ decreases by a factor of a few and~$R$ increases by up to a factor of~$10$. Wave heating causes the brightness of the hydrogen-poor SN progenitors to increase by up to 3 mags in visual bands in the final ten years of evolution. Furthermore, these Ib/c progenitors appear redder in visual bands than predicted by models without wave heating. Our hydrogen-poor wave heated models are approximately consistent with the colors and absolute magnitudes of the progenitors of type Ib SNe iPTF13bvn and SN2019yvr, but low-mass models with a small amount of hydrogen ($\sim \! 10^{-2} \, M_\odot$, with or without wave heating) could also be consistent with those progenitor observations. 

The large expansion of our hydrogen-poor models could also initiate interaction with a close binary companion, which could drive intense pre-supernova mass loss. Future work should study whether this channel is promising for producing more massive CSM. In addition, subsequent work should simulate the mass range~$\sim \! 8\text{--}10\, M_{\odot}$ beyond the propagation of the off-center Ne burning flame and through degenerate silicon ignition, which should be energetic enough to either directly drive a shock \citep{Woosley2015} or excite intense wave heating. Future investigations that are able to model this late-stage nuclear burning will be able to predict whether these stars can eject mass is the final months of their lives.

\section*{Acknowledgments}

We thank Adam Jermyn for supplying the time-dependent convection modules from a MESA development version, and Charlie Kilpatrick for help with HST filter functions. This work was partially supported by NASA grants HSTAR-15021.001-A and 80NSSC18K1017. J.F. acknowledges
support from an Innovator Grant from The Rose Hills
Foundation, and the Sloan Foundation through grant FG2018-10515.
This material is based upon work supported by the National Science Foundation Graduate Research Fellowship under Grant No. DGE‐1745301.

\bibliography{bib,library2}
\end{document}